\documentclass[aps,prx,twocolumn,showpacs,superscriptaddress,longbibliography]{revtex4-2}
\usepackage{amsmath,amssymb,graphicx}
\usepackage{graphicx,epstopdf}
\usepackage{gensymb}
\usepackage[dvipsnames]{xcolor}
\usepackage[T1]{fontenc}
\usepackage[utf8]{inputenc}
\newcommand{\be}{\begin{equation}}
	\newcommand{\ee}{\end{equation}}
\newcommand{\bea}{\begin{eqnarray}}
	\newcommand{\eea}{\end{eqnarray}}
\newcommand{\bse}{\begin{subequations}}
	\newcommand{\ese}{\end{subequations}}

\usepackage{multibib}
\usepackage{color}
\usepackage[colorlinks,bookmarks=false,citecolor=darkblue,linkcolor=red,urlcolor=blue]{hyperref}

\definecolor{darkred}{rgb}{0.7,0.0,0.0}

\definecolor{darkblue}{rgb}{0,0.02,0.45}

\definecolor{darkgreen}{rgb}{0.02,0.45,0.0}

\definecolor{violet}{rgb}{0.8,0.2,0.6}

\begin{document}

\title{Universal fluctuating regime in triangular chromate antiferromagnets}

\author{K. Somesh}
\affiliation{School of Physics, Indian Institute of Science
Education and Research Thiruvananthapuram-695551, India}

\author{Y. Furukawa}
\affiliation{Ames Laboratory and Department of Physics and Astronomy, Iowa State University, Ames, Iowa 50011, USA}

\author{G. Simutis}
\affiliation{Laboratory for Muon Spin Spectroscopy, Paul Scherrer Institut, 5232 Villigen PSI, Switzerland}

\author{F. Bert}
\affiliation{Université Paris-Saclay, CNRS, Laboratoire de Physique des Solides, 91405, Orsay, France}

\author{M. Prinz-Zwick}
\author{N. B\"uttgen}
\affiliation{Experimental Physics V, Center for Electronic Correlations and Magnetism, University of Augsburg, D-86159 Augsburg, Germany}

\author{A. Zorko}
\affiliation{Jo\v{z}ef Stefan Institute, Jamova c. 39, SI-1000 Ljubljana, Slovenia}
\affiliation{Faculty of Mathematics and Physics, University of Ljubljana, SI-1000 Ljubljana, Slovenia}

\author{A. A. Tsirlin}
\email{altsirlin@gmail.com}
\affiliation{Experimental Physics VI, Center for Electronic Correlations and Magnetism, University of Augsburg, 86135 Augsburg, Germany}

\author{P. Mendels}
\email{philippe.mendels@universite-paris-saclay.fr}
\affiliation{Université Paris-Saclay, CNRS, Laboratoire de Physique des Solides, 91405, Orsay, France}

\author{R. Nath}
\email{rnath@iisertvm.ac.in}
\affiliation{School of Physics, Indian Institute of Science Education and Research Thiruvananthapuram-695551, India}
\date{\today}

\begin{abstract}
We report x-ray diffraction, magnetic susceptibility, heat capacity, $^{1}$H nuclear magnetic resonance (NMR), and muon spin relaxation ($\mu$SR) measurements, as well as density-functional band-structure calculations for the frustrated $S=3/2$ triangular lattice Heisenberg antiferromagnet (TLHAF) $\alpha$-HCrO$_{2}$ (trigonal, space group: $R\bar{3}m$).
This compound undergoes a clear magnetic transition at $T_{\rm N} \simeq 22.5$~K, as seen from the drop in the muon paramagnetic fraction and concurrent anomalies in the magnetic susceptibility and specific heat.
Local probes (NMR and $\mu$SR) reveal a broad regime with slow fluctuations down to $0.7\,T_{\rm N}$, this temperature corresponding to the maximum in the $\mu$SR relaxation rate and in the NMR wipe-out. From the comparison with NaCrO$_{2}$ and $\alpha$-KCrO$_{2}$, the fluctuating regime and slow dynamics below $T_{\rm N}$ appear to be hallmarks of the TLHAF with $ABC$ stacking that leads to a frustration of interlayer couplings between the triangular planes. This interlayer frustration is a powerful lever to generate spin states with persistent dynamics and may bear implications to spin-liquid candidates with the triangular geometry.
\end{abstract}

\maketitle

\section{Introduction}
Experimental realization of quantum spin liquid (QSL) remains one of the major challenges in condensed-matter physics~\cite{broholm2020, knolle2019}. An identification of this exotic entangled state in real-world materials could open the way to understanding and utilizing fractionalized excitations, which are particularly interesting in the context of quantum computing~\cite{wen2019}. Theoretical studies established several promising settings for the QSL that were indeed realized~\cite{chamorro2021} in antiferromagnets with the kagome~\cite{mendels2016}, triangular~\cite{Li224004}, and honeycomb~\cite{winter2017} structures, all of them based on the two-dimensional interaction geometries. Relatively less attention has been given to the interlayer couplings that are often neglected in theoretical models but inevitably present in real materials. Moreover, these interlayer couplings are responsible for the formation of three-dimensional long-range magnetic order, so they may be in fact decisive for whether a given material becomes magnetically ordered or develops a fluctuating spin-liquid-like state.

\begin{figure*}
	\includegraphics[height=2.55in, width=7in] {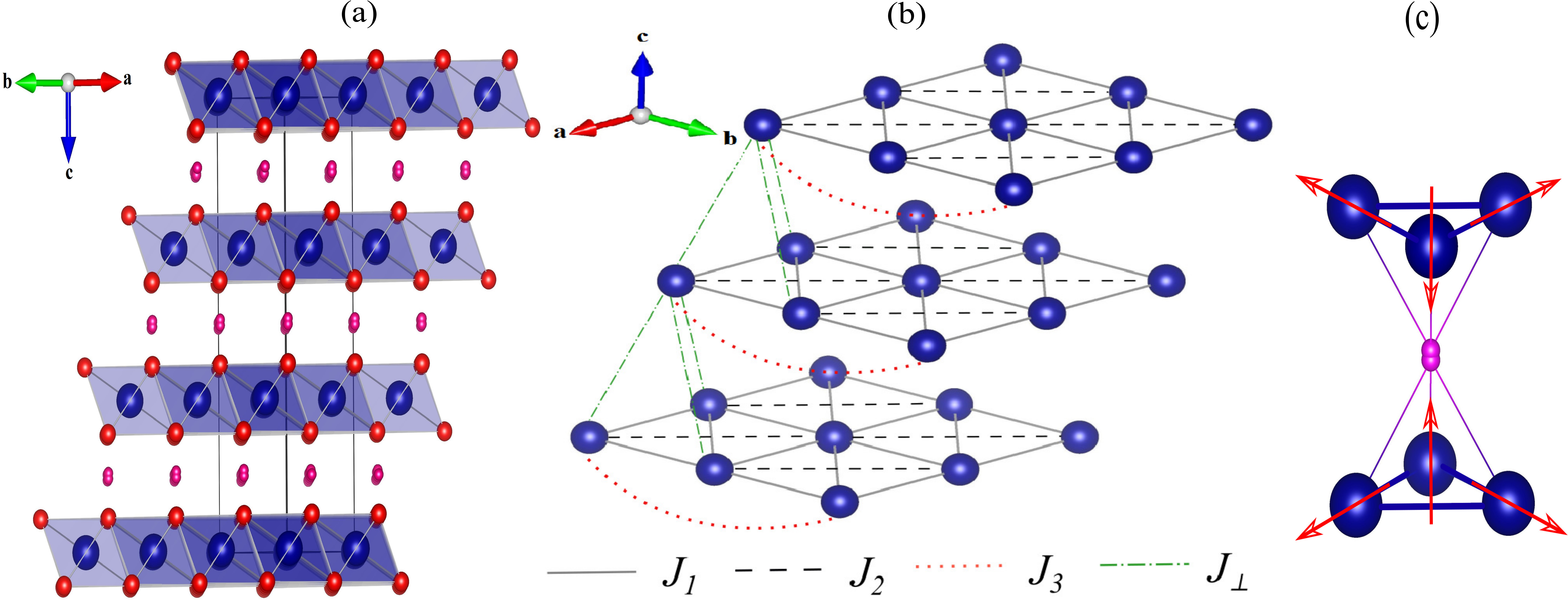}
	\caption{\label{Fig1} (a) Crystal structure of $\alpha$-HCrO${_2}$ consisting of triangular layers formed by CrO${_6}$ octahedra and H atoms at the interstitial sites. (b) Triangular Cr layers showing possible exchange couplings $J_{1}$, $J_{2}$, $J_{3}$, and $J_{\perp}$ between the Cr$^{3+}$ ions. (c) Coupling of H with three Cr$^{3+}$ ions from each layer where spins are arranged in a 120$^\circ$ structure.}
\end{figure*}

Recent work on triangular antiferromagnets~\cite{Li224004} exposed the $A$Yb$X_2$ compound family ($A$ = Na, K and $X$ = O, S, Se) as the largest stock of the QSL candidates known to date. In fact, all members of this family evade long-range magnetic order~\cite{liu2018,baenitz2018,ranjith2019a,ranjith2019b,sarkar2019,bordelon2020}, while some of them additionally show spectral signatures of fractionalized spinon excitations~\cite{dai2021} or excitations of a Dirac QSL~\cite{ding2019,bordelon2019}. One potentially important but hitherto neglected feature of these compounds is the $ABC$-type stacking of their triangular layers that causes frustration of the interlayer couplings, as shown in Fig.~\ref{Fig1}. Each magnetic site is coupled to three sites of the adjacent layer. This arrangement prevents simple ferro- or antiferromagnetic order along $c$.

Here, we explore the effect of this intra and inter-layer frustration using the $A$CrO$_2$ chromates. These compounds are convenient model systems that contain semiclassical spins $\frac32$ and, thanks to the half-filled $t_{2g}$ shell of Cr$^{3+}$, feature isotropic Heisenberg interactions within the triangular planes. Such an interaction regime is known to induce $120^{\circ}$ order in the plane, which is indeed observed in LiCrO$_2$~\cite{Kadowaki6869,Alexander064429,Sugiyama184411,Olariu224401}, but other members of the same family do not show a clear magnetic ordering. In NaCrO$_2$, a transition at $T_{\rm N}\simeq 41$~K is followed by a broad fluctuating regime with slow dynamics extending down to $T^*\simeq 30$\,K~\cite{Olariu167203} for which the occurrence of Berezinskii-Kosterlitz-Thouless (BKT) was suggested~\cite{Hemmida054406}. On the other hand, $\alpha$-HCrO$_2$ with the shortest interlayer distance was recently claimed to evade magnetic order and lie in the vicinity of a spin-liquid phase~\cite{Liu033040}. Using a combination of thermodynamic and local probes applied to $\alpha$-HCrO$_2$, we show that in fact all these chromates -- NaCrO$_2$, $\alpha$-HCrO$_2$, as well as $\alpha$-KCrO$_2$~\cite{Xiao180401} -- develop a very similar phenomenology. They enter a broad universal fluctuating regime below $T_{\rm N}$, followed by the formation of a static and likely incommensurate spin state below $T^*$. Our findings reveal a new mechanism of creating fluctuating spin-liquid-like states that reside on the triangular planes but crucially rely on the interlayer frustration.

\section{Methods}
A polycrystalline sample of $\alpha$-HCrO$_{2}$ was prepared by an hydrothermal method. It involves two steps. First, the intermediate oxide Cr$_8$O$_{21}$ was prepared by heating CrO$_{3}$ (Aldrich, $99.99$\% pure) granules for 12~hours at 250~$^{\circ }$C in air. In the second step, the intermediate product was crushed into fine powder, put into a 1~M LiOH solution in deionized water, heated at 200~$^{\circ }$C in a teflon-lined stainless steel autoclave for about one week, and subsequently furnace-cooled. The phase purity of the product was confirmed by powder x-ray diffraction (XRD) using a PANalytical x-ray diffractometer (Cu \textit{K$_{\alpha}$} radiation, $\lambda_{av}=1.54182$ \AA). The temperature-dependent powder XRD measurement was performed over the temperature range 15~K$\leq T \leq 300$~K using the low-temperature attachment (Oxford Phenix) to the diffractometer.

Magnetization ($M$) measurements were performed as a function of temperature ($T$) and applied field ($H$) using a superconducting quantum interference device (MPMS-3, Quantum Design). Heat capacity [$C_{\rm p}(T)$] as a function of $T$ and $H$ was measured on a small piece of sintered pellet using the relaxation technique in the physical property measurement system (PPMS, Quantum Design).

The NMR measurements were carried out using pulsed NMR techniques on $^{1}$H (nuclear spin $I=1/2$ and gyromagnetic ratio $\gamma _{\rm N}/2\pi = 42.575$~MHz/T) nuclei in the temperature range 1.6~K~$\leq T\leq 250$~K and at different radio frequencies. The spectra were obtained either by Fourier transform (FT) of the NMR echo signal or by sweeping the magnetic field.
The NMR shift $K=(\nu -\nu _{\rm ref})/\nu _{\rm ref}$ was determined by measuring the resonance frequency of the sample ($\nu $) with respect to nonmagnetic reference H$_{2}$O (resonance frequency $\nu _{\rm ref}$).
The $^{1}$H spin-lattice relaxation rate ($1/T_{1}$) was measured by the conventional inversion recovery method. The spin-spin relaxation rate ($1/T_{2}$) was measured through the decay of the echo integral with variable spacing between the $\pi/2$ and $\pi$ pulses.

The muon spin rotation/relaxation ($\mu$SR) experiments were carried out at the $\pi$M$_3$ beam line using the GPS spectrometer at the Paul Scherrer Institute (PSI), Switzerland~\cite{Amato093301}. The zero-field (ZF) and the weak transverse-field (wTF) $\mu$SR measurements were performed at temperatures ranging from $\sim 1.5$~K to 50~K. The 100\% spin-polarized $\mu^+$ (spin-1/2 and gyromagnetic ratio $\gamma_{\mu} = 135.5$~MHz/T) with the momentum of $\sim 28.6$~MeV/c were implanted into the sample. Because of their positive charge, muons come at rest in well-defined sites where their electrostatic energy is minimized, i.e. in oxides typically 1~{\AA} away from an oxygen O$^{2-}$ site. They usually interact with surrounding moments through a dipolar coupling. In a paramagnetic state, the electronic moments fluctuate
fast ($\sim 10^{-12}$~s) on the $\mu$SR time scale and the static field sensed by the muons has only a nuclear origin with a typical value of a few Gauss, while dynamical electronic fields lead to motional narrowing. On the contrary, when electronic moments slow down, the evolution of the ZF polarization with time witnesses both their dynamics and freezing. At the base temperature ($T \ll T_{\rm N}$), very high counting statistics ($\sim 115$ Mevents) were taken in order to track the fast-decaying polarization measured in zero-field, where the contribution from static electronic moments dominates.

Density-functional (DFT) band-structure calculations were performed in the \texttt{FPLO} code~\cite{fplo} using the Perdew-Burke-Ernzerhof flavor of the exchange-correlation potential~\cite{Perdew3865}. Correlation effects in the Cr $3d$ shell were taken into account of the mean-field DFT+$U$ level using the on-site Coulomb repulsion $U_d=2$\,eV, Hund's coupling $J_d=1$\,eV, and atomic limit for the double-counting correction~\cite{Janson064417,Janson214424}. Experimental structural parameters for LiCrO$_2$~\cite{LuA1454}, NaCrO$_2$~\cite{Scheld151}, and $\alpha$-KCrO$_2$~\cite{Scheld151} have been used. In the case of $\alpha$-HCrO$_2$, the position of hydrogen remains somewhat uncertain. We thus considered two limiting cases: i) hydrogen placed in the middle between the CrO$_2$ layers, with the $R\bar 3m$ symmetry preserved; ii) hydrogen displaced toward one of the layers. In the latter case, the symmetry is reduced to $R3m$, and the individual CrO$_6$ octahedra are slightly distorted, because the Cr--O distances can be shorter or longer depending on how strongly the oxygen atom binds to the hydrogen. Room-temperature atomic positions from Ref.~\cite{Ichikawa1875} have been used, as they feature the largest distortion and allow one to test its effect on magnetism. Exchange parameters $J_{ij}$ and single-ion anisotropy $D$ enter the spin Hamiltonian,
\begin{equation}
	\mathcal H=\sum_{\langle ij\rangle} J_{ij}\mathbf S_i\mathbf S_j -\sum_i D(S_i^z)^2,
	\label{eq1}
\end{equation}
where the summation is over atomic pairs $\langle ij\rangle$, and $S=\frac32$. The parameters of this Hamiltonian were obtained by a mapping procedure~\cite{Xiang224429} from total energies of magnetically ordered states.

\section{Results}
\subsection{X-ray Diffraction}
\begin{figure}
	\includegraphics[width=\columnwidth] {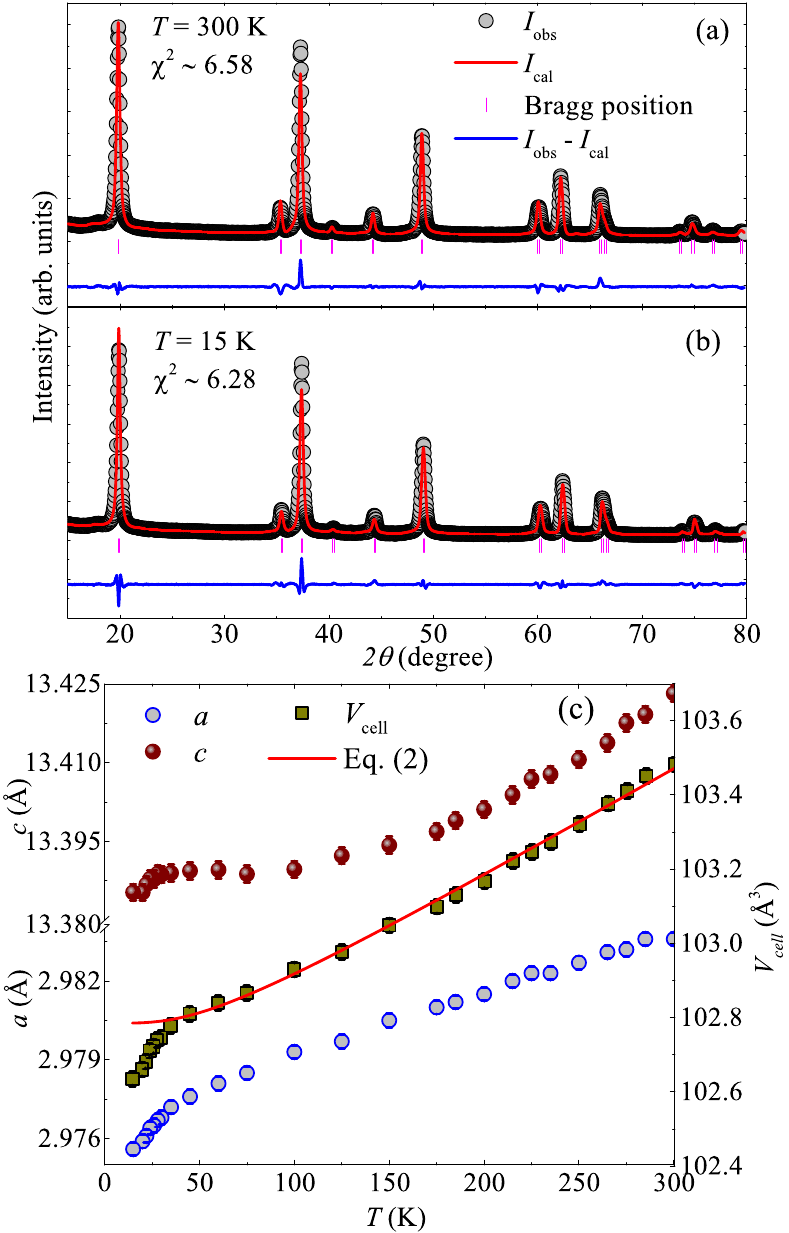}
	\caption{\label{Fig2} Powder XRD data measured at (a) $T = 300$ and (b) $T = 15$~K. The red solid line represents the Rietveld fit of the data. The Bragg positions are indicated by pink vertical bars and the solid blue line at the bottom denotes the difference between the experimental and calculated intensities. (c) The variation of the lattice parameters ($a$, $c$, and $V_{\rm cell}$) as a function of temperature. The solid line denotes the fit of the $V_{\rm cell}(T)$ by Eq.~\eqref{eq2}.}
\end{figure}
In order to confirm the phase purity and to check the presence of any structural distortions, powder XRD data were collected at various temperatures. The Rietveld refinements of the XRD patterns were executed using the \verb"FullProf" software package~\cite{Juan55} with the initial parameters taken from Ref.~\cite{Ichikawa1875}. Figure~\ref{Fig2} presents the powder XRD patterns at $300$ and $15$~K along with the Rietveld fits. At room temperature, all the peaks could be indexed based on the space group $R\bar{3}m$ (No.~166) suggesting phase purity of the sample. The obtained lattice parameters at room temperature are $a = b = 2.9836(1)$~\AA, $c = 13.4232(1)$~\AA, and unit cell volume $V_{\rm cell} \simeq 103.48$~\AA$^3$, which are comparable with the previous report~\cite{Ichikawa1875}. No extra features or peaks were observed in the XRD scans implying the absence of any structural transitions down to 15~K. On the other hand, around 30~K both lattice parameters ($a$ and $c$) clearly deviate from their anticipated low-temperature behavior [see Fig.~\ref{Fig2}(c)]. Such weak kinks, humps, and changes of slope are commonly observed in magnetic compounds in the vicinity of their magnetic transitions, where lattice symmetry does not change but a magnetoelastic coupling leads to anomalies in thermal expansion~\cite{chatterji2012,reschke2020}.

The temperature variation of the lattice constants ($a$ and $c$) and unit cell volume ($V_{\rm cell}$) as shown in Fig.~\ref{Fig2}(c) are found to decrease monotonically upon cooling down to $T \approx 45$~K.
$V_{\rm cell}(T)$ above 45~K was fitted by the equation~\cite{Islam174432}
\begin{equation}
V(T)=\gamma U(T)/K_0+V_0,
\label{eq2}
\end{equation}
where $V_0$ is the cell volume in the zero temperature limit, $K_0$ is the bulk modulus, and $\gamma$ is the Gr$\ddot{\rm u}$neisen parameter. $U(T)$ is the internal energy, which can be derived in terms of the Debye approximation as
\begin{equation}
U(T)=9nk_{\rm B}T\left(\frac{T}{\theta_{\rm D}}\right)^3\int_{0}^{\theta_{\rm D}/T}\dfrac{x^3}{e^x-1}dx.
\label{eq3}
\end{equation}
Here, $n$ is the number of atoms in the unit cell and $k_{\rm B}$ is the Boltzmann constant. Using this approximation [see Fig.~\ref{Fig2}(c)], the Debye temperature ($\theta_{\rm D}$) and other parameters were estimated to be $\theta_{\rm D} \simeq 195$~K, $\gamma/K_0 \simeq 2.95 \times 10^{-5}$~Pa$^{-1}$, and $V_0 \simeq 102.784$~\AA$^{3}$.

\subsection{Magnetization}
\begin{figure}
	\includegraphics[width=\columnwidth] {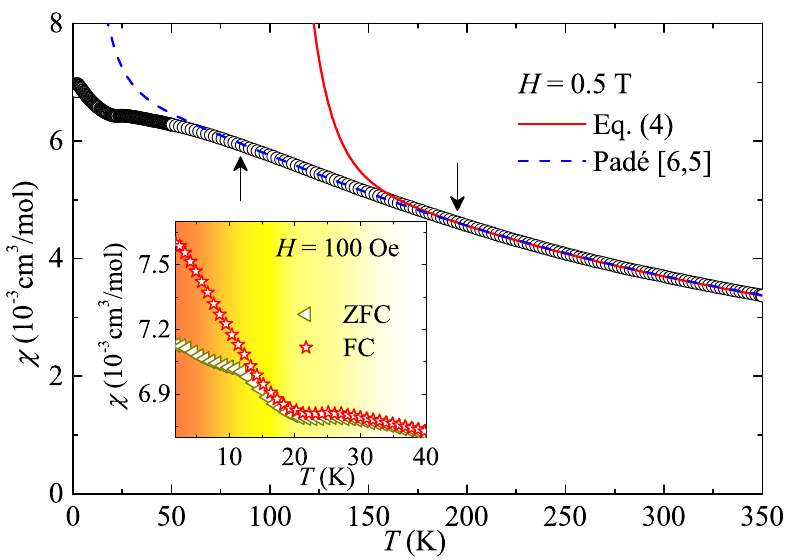}
	\caption{\label{Fig3} Temperature dependence of magnetic susceptibility [$\chi(T)$] measured in an applied field of $0.5$~T and in the temperature range 1.8~K$\leq T \leq 350$~K. The solid and dashed lines denote the fits by Eq.~\eqref{eq4} and Pad\'{e} approximation, respectively. Arrows signal the limit of validity of HTSE or Pad\'{e} approximation. Inset: Zero-field-cooled (ZFC) and field-cooled (FC) magnetic susceptibilities as a function of temperature measured in an field of $100$~Oe.}
\end{figure}
Previous magnetic susceptibility $\chi(T)$ and $^{1}$H NMR measurements were restricted to temperatures above the possible magnetic ordering~\cite{Meisenheimer831,Ibers1620}. An antiferromagnetic transition at $T_{\rm N} \sim 25$~K has been reported from the heat capacity measurements~\cite{Matsuo129}.

The temperature-dependent magnetic susceptibility $\chi(T)$ ($\equiv M/H$) measured in an applied field of $0.5$~T is shown in Fig.~\ref{Fig3}. As expected in the paramagnetic regime, $\chi(T)$ increases with decreasing temperature in a Curie-Weiss manner but becomes flatter toward lower temperatures when spin-spin correlations set in. A hump around 27\,K coincides with the lattice anomaly discussed above. At even lower temperatures, $\chi(T)$ shows an upturn around 20\,K and a splitting at 12\,K between field-cooled (FC) and zero-field-cooled (ZFC) curves measured in a weak applied field. While the latter signature may be indicative of spin freezing, our frequency-dependent AC susceptibility measurement (see, Supplementary Materials) rules out the possibility of a conventional spin-glass transition.


Traditionally, $\chi (T)$ is fitted by the sum of a temperature independent term ($\chi_0$) and of a Curie-Weiss law, $\chi_0+\frac{C}{T-\theta_{\rm CW}}$ in order to obtain the Curie constant $C$ and the characteristic CW temperature $\theta_{\rm CW}$. $C$ yields the effective moment while $\theta_{\rm CW}$ represents the energy scale of the total exchange interactions. $\theta_{\rm CW}$ is given by: $|\theta_{\rm CW}|=\frac{JzS(S+1)}{3k_{\rm B}}$, where $z=6$ is the number of nearest-neighbours of Cr$^{3+}$ ions and $J~(=J_1)$ is the intra-layer exchange coupling~\cite{Domb1964}. This requires the $T$ range of the measurements to fall into the high-temperature regime, $T\gg|\theta_{\rm CW}|$ which is not the case here. Indeed, our $\chi(T)$ data are limited up to 350~K and $\theta_{\rm CW}$ for HCrO$_2$ is in the range $220 - 270$~K~\cite{Meisenheimer831}.

We therefore used high-temperature series expansion (HTSE), now available up to 11$^{\rm th}$ order~\cite{Schmidt104443} and Pad\'{e} approximants~\cite{Lohmann014415} to fit our data,
\begin{equation}
\chi=\chi_{0}+\chi_{\rm spin}(T)
\label{eq4}
\end{equation}
and
\begin{eqnarray}
	\begin{split}
		\lefteqn{\chi_{\rm spin}(T) = \frac{N_{\rm A}g^2\mu_{B}^{2}}{k_{\rm B}T} \times \Bigg[\frac{5}{4}-\frac{75}{8}x+\frac{225}{4}x^2}\\&
		-\frac{18765}{64}x^3+\frac{712175}{512}x^4-\frac{6328661}{1024}x^5+\frac{643219519}{24576}x^6\\&
		-\frac{36677316665}{344064}x^7+\frac{1154751891527}{27525127}x^8\\&
		-\frac{1888217340683}{1179648}x^9+\frac{11768319087087113}{1981808640}x^{10}\Bigg],
	\end{split}
	\label{eq5}
\end{eqnarray}
with $x = \frac{J}{k_{\rm B}T}$. Figure~\ref{Fig3} depicts the fitting of $\chi(T)$ data in the validity $T$-range of Eq.~\eqref{eq4}, $k_{\rm B}T > 8J$, corresponding to $T> 195$~K. Fixing $g=2$ from the ESR data of Ref.~\cite{Hemmida054406}, we obtain $J/k_{\rm B} = (24 \pm 0.2)$~K and $\chi_{0} = -(5.35 \pm 0.45) \times 10^{-5}$~cm$^{3}$/mol. An extension of the validity of the fit down to $T\sim 3.5 J$ can be obtained using Pad\'{e} approximants which are now available. The [6,5] Pad\'{e} extension of our fit is shown as a dashed blue line in Fig.~\ref{Fig3} and perfectly matches with our data in its validity domain, $T \gtrsim 85$~K, with $J/k_{\rm B} = 24$~K and $\chi_0 = -5.35 \times 10^{-5}$~cm$^3$/mol.


\subsection{Heat Capacity}
\begin{figure}
	\includegraphics[width=\columnwidth] {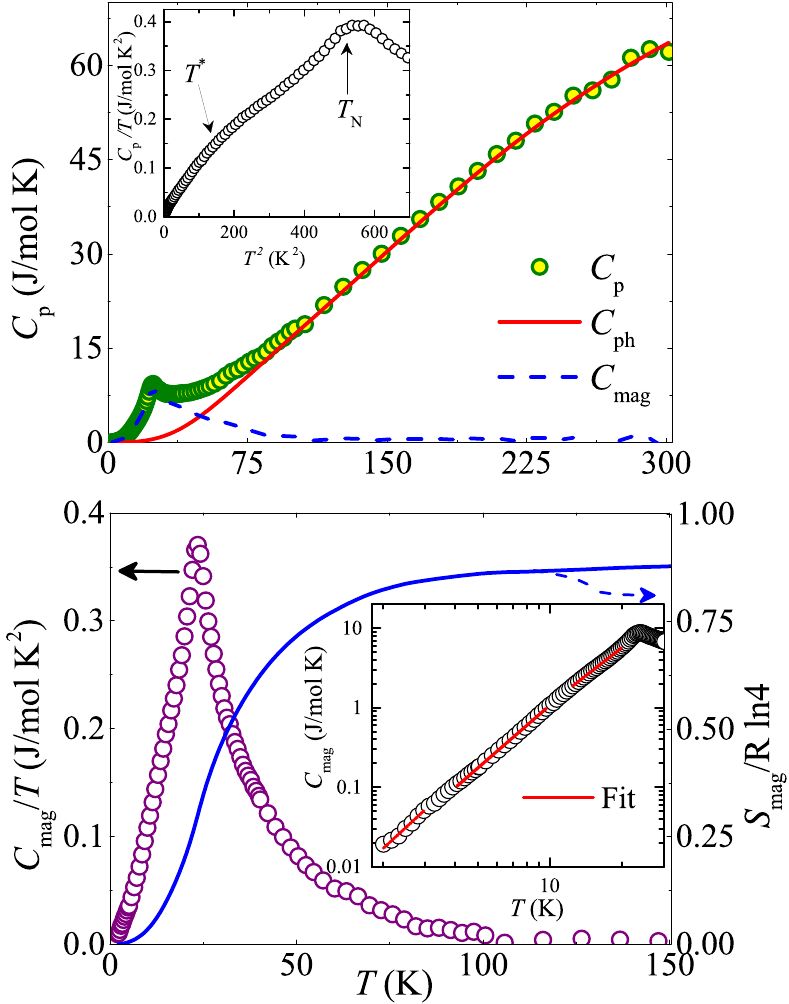}
	\caption{\label{Fig4} Upper panel: Temperature-dependent heat capacity $C_{\rm p}(T)$ of $\alpha$-HCrO$_{2}$ measured in zero applied magnetic field. The solid line represents the simulated phonon contribution $C_{\rm ph}(T)$ and the dashed line represents the magnetic contribution $C_{\rm mag}(T)$. Inset: Plot of $C_{\rm p}(T)/T$ vs $T^2$ below $T_{\rm N}$, highlighting the shoulder at $T^* \simeq 12$~K. Lower panel: $C_{\rm mag}/T$ and $S_{\rm mag}/R\ln 4$ in the left and right $y$-axes, respectively, are plotted as a function of temperature. Inset: Logarithmic plot of $C_{\rm mag}$ vs $T$, with solid lines showing the fits with a power-law, $C_{\rm mag} = aT^{\alpha}$, in different temperature regimes.}
\end{figure}
The heat capacity ($C_{\rm p}$) data are shown in Fig.~\ref{Fig4} (upper panel) as a function of temperature measured in zero magnetic field. At high temperatures, $C_{\rm p}(T)$ is entirely dominated by the contribution of phonon excitations ($C_{\rm ph}$) and the value of $C_{\rm p}$ at $300$~K is about $\sim 62$~J/mol K. This value is close to the expected Dulong-Petit lattice heat capacity of $C_{\rm v}=3nR \simeq 74.8$~J/mol~K, where $n$ is the number of atoms per formula unit~\cite{Fitzgerel545}. In $\alpha$-HCrO$_{2}$, H is the lightest element expected to have a very high Debye frequency compared to other elements, and therefore we chose $n = 3$ for the calculation. At around $24$~K, the heat capacity shows a clear and very broad anomaly that confirms the magnetic transition. With further decrease in temperature, $C_{\rm p}(T)$ decreases gradually towards zero. At low temperatures, $C_{\rm p}(T)$ is dominated by the magnetic contribution $C_{\rm mag}$.

In order to estimate the phonon part of the heat capacity, the $C_{\rm p}(T)$ data were fitted above 100~K by a sum of Debye contributions
\begin{equation}
\label{eq8}
C_{\rm p}(T) = 9R\displaystyle\sum\limits_{n=1}^{3} c_n \left(\frac{T}{\theta_{Dn}}\right)^3 \int_0^{\frac{\theta_{Dn}}{T}} \frac{x^4e^x}{(e^x-1)^2} dx.
\end{equation}
Here, $R$ is the molar gas constant, $\theta_{Dn}$ are the characteristic Debye temperatures, and $c_n$ are the integer coefficients indicating the contributions of different atoms to $C_{\rm p}(T)$. A similar approach has been chosen previously to estimate the phonon contribution in different types of frustrated magnets~\cite{Ahmed214413}. Figure.~\ref{Fig4} (upper panel) presents the fit of $C_{\rm p}(T)$ by Eq.~\eqref{eq8} with $c_1 = 1$, $c_2 = 1$, and $c_3 = 2$. Here, $c_1$, $c_2$, and $c_3$ represent the number of H, Cr, and O atoms, respectively. The sum of $c_n$ is thus equal to $4$, the number of atoms per formula unit. We have used three different Debye temperatures: $\theta_{D1}$ for H$^{1+}$, $\theta_{D2}$ for Cr$^{3+}$, and $\theta_{D3}$ for O$^{2-}$. Finally, the high-$T$ fit was extrapolated down to $2$~K and $C_{\rm {mag}}(T)$ was estimated by subtracting $C_{\rm {ph}}(T)$ from $C_{\rm {p}}(T)$ [see Fig.~\ref{Fig4} (upper panel)].

$C_{\rm {mag}}(T)/T$ is plotted as a function of temperature in the lower panel of Fig.~\ref{Fig4}. It extends up to 100~K, even though the leading exchange coupling $J$ is about 24~K only. For the validation of the fitting procedure, we calculated the total magnetic entropy ($S_{\rm mag}$) by integrating $C_{\rm mag}(T)/T$ between $2$~K and high-temperatures as
$S_{\rm{mag}}(T) = \int_{\rm 2\,K}^{T}\frac{C_{\rm {mag}}(T')}{T'}dT'$. The obtained magnetic entropy at 150~K is $S_{\rm{mag}}\simeq$ $10.4$~J/mol K. This value of $S_{\rm mag}$ corresponds to $\sim 90\%$ (see the lower panel of Fig.~\ref{Fig4}) of the expected theoretical value $S_{\rm mag} = R\ln(2S+1) = 11.5$~J/mol K for a $S = 3/2$ system. Moreover, the entropy at $T_{\rm N}$ is found to be only one-third of the total entropy and the remaining entropy is distributed above $T_{\rm N}$ due to short-range (in-plane) spin correlations. This is in contrast with the conventional LRO where the entropy is recovered completely just above $T_{\rm N}$, and confirms the strongly frustrated nature of $\alpha$-HCrO$_2$~\cite{Alexander064429,Takatsu104424}.

The broad maximum in $C_{\rm mag}$ around 24~K confirms the magnetic transition, whereas a broad shoulder in $C_{\rm p}/T$ vs $T^2$ (inset of the upper panel of Fig.~\ref{Fig4}) suggests another magnetic instability at $T^{*} \simeq 12$~K that coincides with the temperature at which the ZFC and FC susceptibilities bifurcate. This feature is quite similar to that reported for CuCrO$_2$ and $\alpha$-KCrO$_2$ previously~\cite{Okuda134423,Xiao180401}.


The logarithmic plot of $C_{\rm mag}$ vs $T$ below $T_{\rm N}$ (see, lower inset of Fig.~\ref{Fig4}) reveals a nearly linear behavior, although a closer inspection of the data suggests that different power-law exponents are obtained in different temperature intervals. Using $C_{\rm mag} = aT^{\alpha}$, we find $a \simeq 0.0052$~J-mol$^{-1}$-K$^{-4}$ and $\alpha \simeq 2.2$ for 12.5~K$\leq T \leq 20$~K, $a \simeq 0.0026$~J-mol$^{-1}$-K$^{-4}$ and $\alpha \simeq 2.6$ for 4~K$\leq T \leq 10$~K, and $a \simeq 0.0025$~J-mol$^{-1}$-K$^{-4}$ and $\alpha \simeq 2.8$ for 2~K$\leq T \leq 3$~K. Considering $\alpha=2$ and $\alpha=3$ expected for antiferromagnets in 2D and 3D, respectively, we conclude that spin correlations gradually evolve from a 2D behavior immediately below $T_{\rm N}$ toward a 3D behavior well below $T^*$.

\subsection{$^{1}$H NMR}
   \begin{figure}
	\includegraphics[height=5in, width=3.4in] {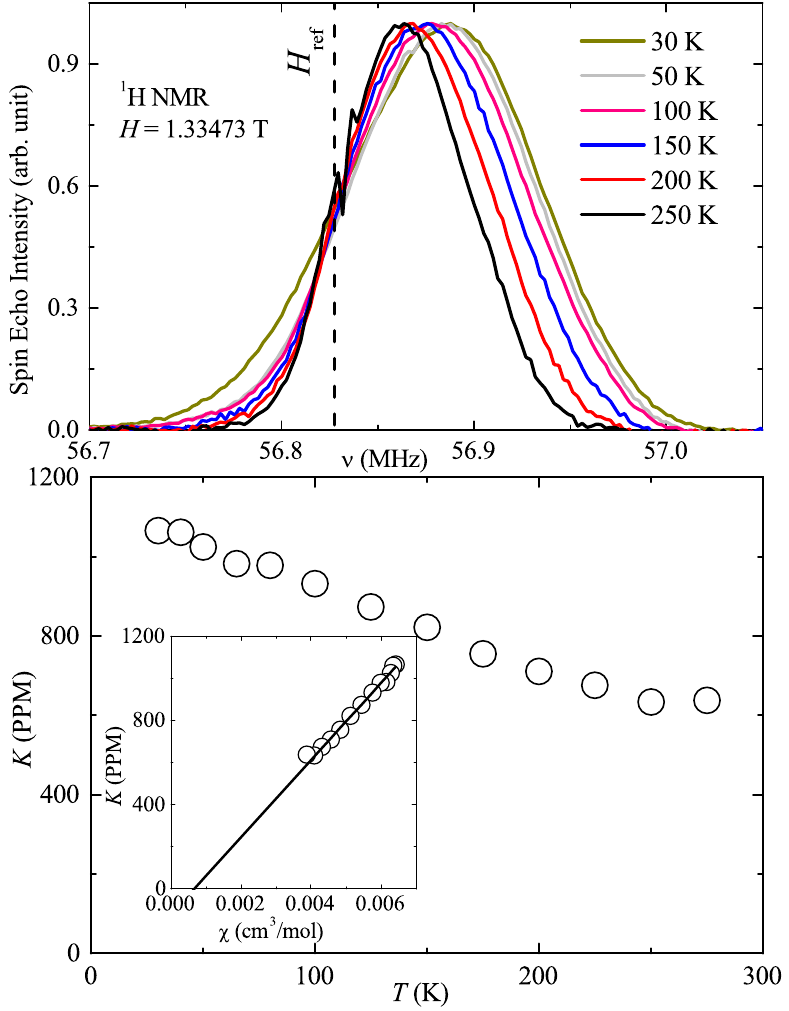}
	\caption{\label{Fig5} Upper panel: FT $^{1}$H NMR spectra at different temperatures for the polycrystalline $\alpha$-HCrO$_{2}$ sample measured at $H = 1.33473$~T. The vertical dashed line corresponds to the $^{1}$H resonance frequency of the non-magnetic reference sample. Lower panel: $^{1}$H NMR shift $K$ as a function of $T$. Inset: $^{1}$H NMR shift vs $\chi$ with temperature as an implicit parameter. The solid line represents the linear fit.}
\end{figure}
The crystal structure of $\alpha$-HCrO$_2$ [Fig.~\ref{Fig1}(c)] features a single crystallographic H site located in between two Cr-triangles. The $^{1}$H nucleus is then coupled to three Cr$^{3+}$ ions from each of the two adjacent layers. Thus, one can probe static and dynamic properties of the Cr$^{3+}$ spins by performing $^{1}$H NMR. We have measured $^{1}$H NMR spectra by doing a Fourier Transform of the echo signal at different temperatures, keeping the field persistent. Figure~\ref{Fig5} (upper panel) presents the FT $^{1}$H NMR spectra at different temperatures. We indeed observed a single spectral line, as expected for a $I=1/2$ nucleus. The line position was found to increase weakly with decreasing temperature. The lower panel of Fig.~\ref{Fig5} presents the temperature variation of the NMR shift ($K$) for $T > 30$~K. The slope of the linear fit using $K(T)=K_{0}+\frac{A_{\rm hf}}{N_{\rm A}} \chi_{\rm spin}(T)$ (where $K_{0}$ is the temperature-independent chemical shift) of the $K$ vs $\chi$ plot yields the hyperfine coupling constant $A_{\rm hf} \simeq 1024$~Oe/$\mu_{\rm B}$ between the $^{1}$H nucleus and the Cr$^{3+}$ electronic spins (see, inset of Fig.~\ref{Fig5}).
This value of $A_{\rm hf}$ is almost six times smaller than $^{7}$Li in LiCrO$_2$~\cite{Alexander064429,Olariu167203} but much larger than the expected dipolar coupling for $\alpha$-HCrO$_2$~\footnote{The dipolar coupling constant at the H site is calculated using lattice sum with an assumption of $1\mu_{\rm B}$ magnetic moment on each Cr$^{3+}$ ion to be $\sim 0.150$~kOe/$\mu_{\rm B}$ per Cr$^{3+}$ ion.}. The latter also indicates a substantial overlap of the hydrogen $1s$ orbital with the $3d$ orbitals of Cr$^{3+}$ ion via $2p$ orbitals of O. This further explains why the inter-layer exchange coupling through H is significant, as estimated in Sec.~F.

\begin{figure}
	\includegraphics[width=\columnwidth] {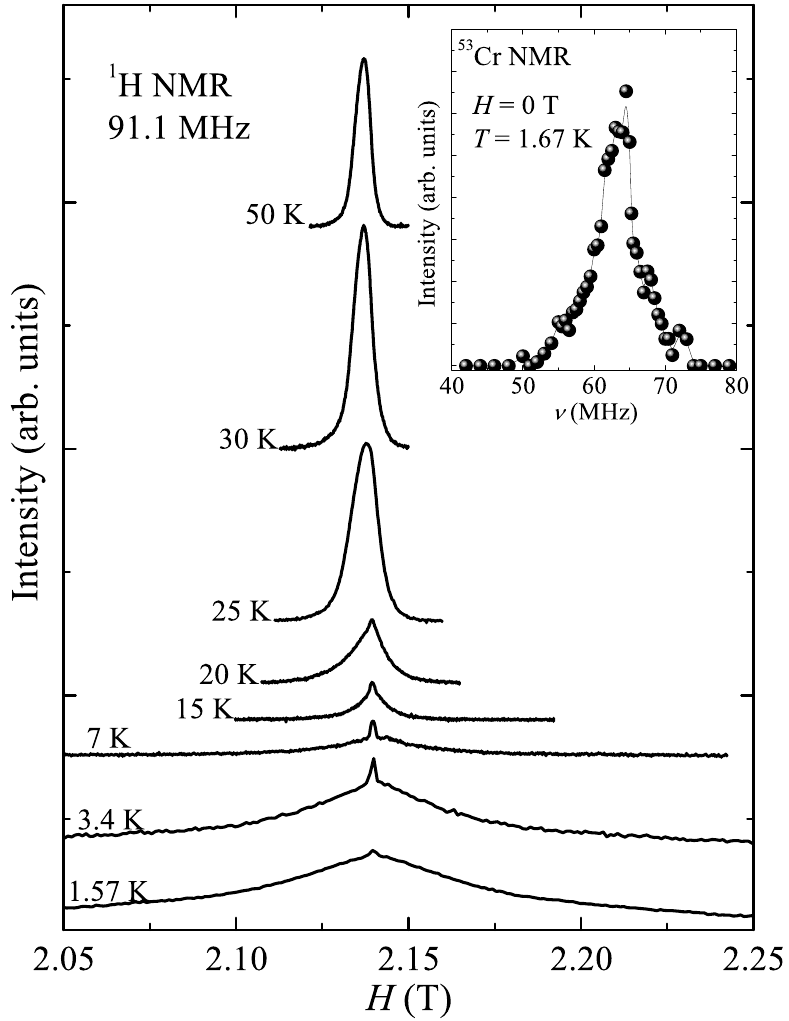}
	\caption{\label{Fig6} Field-sweep $^{1}$H NMR spectra at different temperatures around $T_{\rm N}$ measured on the polycrystalline $\alpha$-HCrO$_{2}$ sample at $91.1$~MHz. Inset: $^{31}$Cr NMR spectrum at $T = 1.67$~K measured in zero field.}
\end{figure}

\begin{figure}
	\includegraphics[width=\columnwidth] {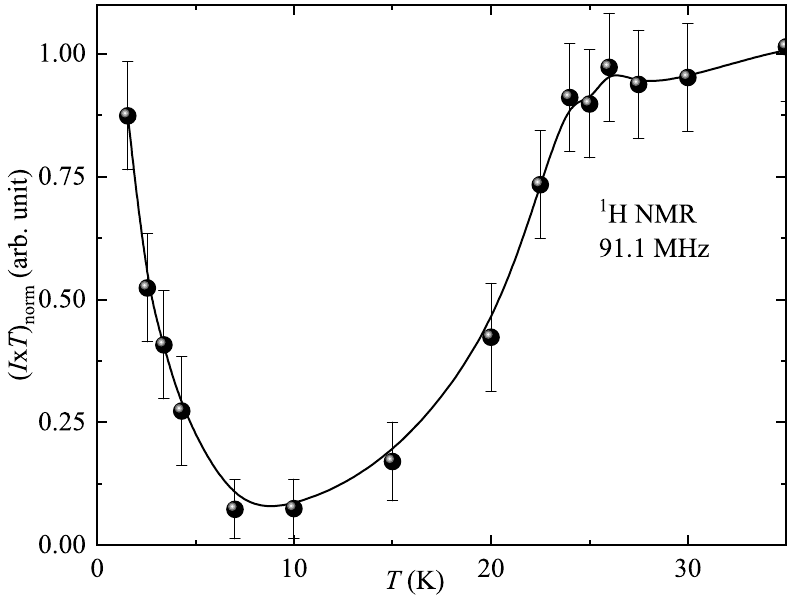}
	\caption{\label{Fig7} Integrated NMR intensity as a function of temperature.}
\end{figure}
Figure~\ref{Fig6} shows the field-sweep NMR spectra around and below $T_{\rm N}$. As the temperature approaches $T_{\rm N}$ from above, the NMR signal intensity decreases. Below $T_{\rm N}$, the signal is reduced drastically, whereas the line width does not show any significant increase. The loss of the NMR signal continues well below $T_{\rm N}$ but never becomes complete, and below $\sim 10$~K the intensity quickly recovers. Below 10~K, the NMR line broadens and develops a triangular shape which is independent of the applied field. This broadening is indicative of local magnetic fields appearing at the $^1$H site, but it is remarkable that such a broadening appears only below 10~K and not immediately below $T_{\rm N}$, as in conventional 3D antiferromagnets. This is a clear indication that the NMR signal is wiped out over a broad temperature window, below $T_{\rm N}$ and a pure static state appears only below 10~K. The wipe out effect is very well evident in the temperature dependent integrated NMR intensity plot in Fig.~\ref{Fig7}. A rectangular line shape on the polycrystalline sample is expected in an ordered collinear antiferromagnet~\cite{Kikuchi2660,Nath024431}. The triangular line shape observed at low temperatures in $\alpha$-HCrO$_2$ is reminiscent of an incommensurate magnetic order~\cite{Kontani672,Ranjith014415} that may appear in triangular antiferromagnets upon a distortion of the commensurate $120^{\circ}$ state. Further, as the intrinsic signal is lost below $T_{\rm N}$ a very narrow central line becomes prominent at the zero shift position, on top of the broad spectrum. This narrow line persists down to the lowest measured temperature and can be attributed to the effect of defects and/or a small amount of non-magnetic impurities.

We have also measured the $^{53}$Cr NMR ($I=3/2$ and $\gamma _{\rm N}/2\pi = 2.40094$~MHz/T) spectrum by sweeping the frequency in zero field at $T = 1.67$~K (see, inset of Fig.~\ref{Fig6}). The observation of $^{53}$Cr zero field NMR signal is a direct evidence of a static magnetic ordering of Cr$^{3+}$ moments in $\alpha$-HCrO$_2$ well below $T^*$. From the peak position, the internal field at the Cr site is estimated to be $|H_{\rm int}| \sim 26.4$~T. Such a large value of $H_{\rm int}$ is comparable to the reported value $\sim 28.64$~T for YCrO$_3$ and $\sim 27.0$~T for CuCrO$_2$~\cite{Takeda165107,*Smolnikov674}.

\subsection{$\mu$SR}
In order to probe the dynamics in the fluctuating regime in more detail, we performed complementary $\mu$SR experiments.
Owing to its much shorter time window (10~ns~-~15~$\mu$s), $\mu$SR is better suited than NMR in tracking the persisting dynamics in slowly fluctuating magnets. In addition, $\mu$SR allows us to probe \emph{all} sites, whereas only a weak fraction of the sites were detected in NMR between 20 and 3~K due to the wipeout effect.
\begin{figure}
	\includegraphics[width=\columnwidth]{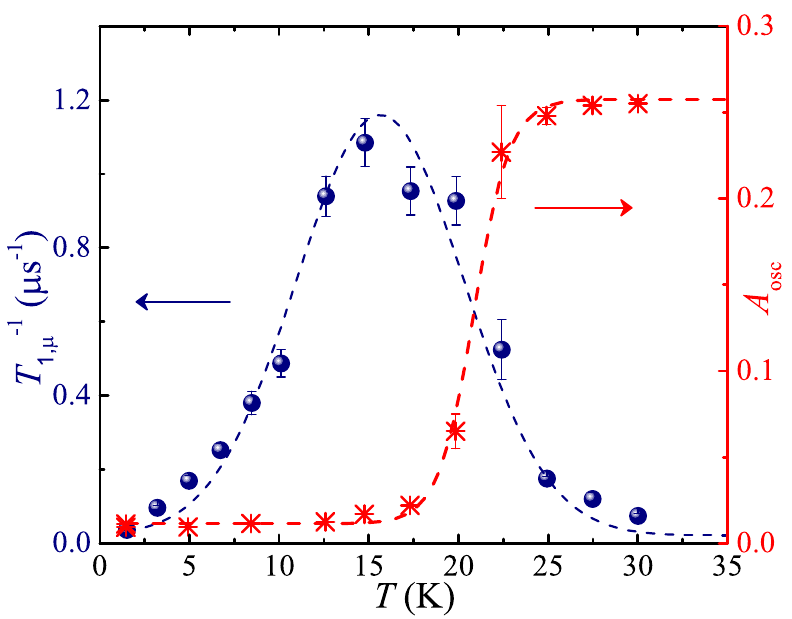}
	\caption{\label{wTF} Left: $T$-variation of the muon relaxation rate; the dashed line is a guide to the eyes. Right: weak transverse field $\mu$SR asymmetry (see text). The dashed line is a fit to a phenomenological broadened "step" function $(A_{\rm tot} - A_{\rm b})/[1+e^{((T_{\rm N} -T)/\Delta T_{\rm N})}]+A_{\rm b}$.}
\end{figure}

$\mu$SR experiments under a weak applied field perpendicular to the initial muon spin polarization (wTF), were performed in order to track the spin freezing and the static magnetic volume fraction versus temperature. Indeed, only muons stopping close to paramagnetic (unfrozen) sites precess around the applied field direction and produce long-lived oscillations of the polarization reflected in the measured asymmetry. On the contrary, muons close to magnetically frozen parts experience much larger fields with a large distribution and yield a strongly damped contribution in the 0.1~$\mu$s range to the polarization as is evident in zero-field experiments presented next.

The wTF asymmetry was fitted using a standard function for $t > 0.5~\mu$s
\begin{equation}
	A_{\rm wTF}(t) = A_{\rm osc} \cos(\gamma_{\mu} B_{\rm ext} t+ \phi) e^{-(\sigma^2 t^2/2)}+ A_{\rm tail}e^{-\lambda t},
\end{equation}
where $A_{\rm osc}$ is the wTF oscillating asymmetry and $A_{\rm tail}$ is introduced to account for the non-oscillating long-time relaxing frozen part (1/3$^{\rm rd}$ tail). The fitted values for $A_{\rm osc}$ are shown in Fig.~\ref{wTF} (right). An abrupt loss of asymmetry is observed at a temperature of 22.5~K - taken as the most precise definition of $T_{\rm N}$, within a $\Delta T_{\rm N} \sim 3$~K range around $T_{\rm N}$, indicating a uniform freezing in the sample. At lower temperatures, a constant $A_b \sim 0.01$ asymmetry is found which corresponds to $\sim 4\%$ muons sitting in a non magnetic part of the sample, possibly defects or an impurity phase. Note that this is in-line with the spectral weight of the persisting narrow peak found in the NMR experiment.

\begin{figure}
\includegraphics[width=\columnwidth]{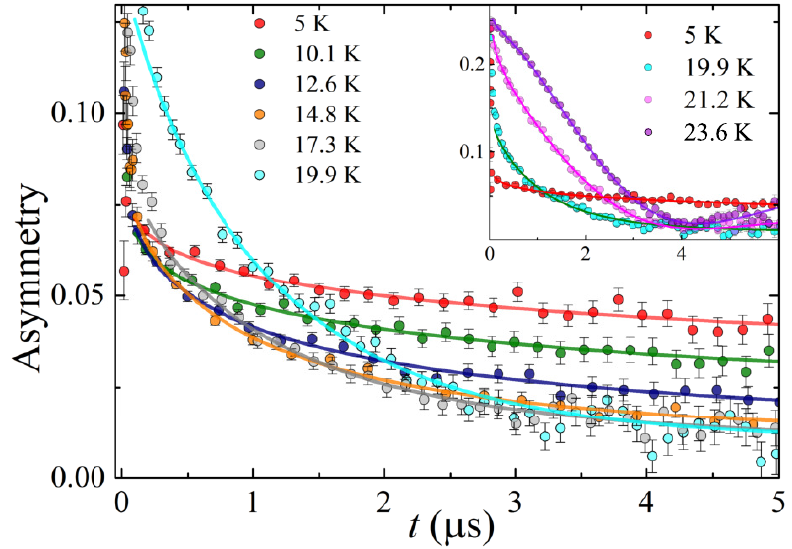}
\caption{\label{ZFmuSR} $\mu$SR asymmetry in zero external field at various temperatures. The solid lines are fits described in the text.}
\end{figure}

Typical zero-field asymmetry curves are presented in Fig.~\ref{ZFmuSR}. Above $T_{\rm N}$, in the fast fluctuation regime, muons mainly sense a weak $\sim$~4.6~G static nuclear magnetic field mainly originating from H nuclei. The decrease of the $\mu^+$ polarization has the expected Kubo-Toyabe shape, Gaussian at early times and modulated by a slowly exponential relaxing envelope associated with electronic spin fluctuations.
When decreasing the temperature below $T_{\rm N}$, similar to the NaCrO$_2$ case~\cite{Olariu167203}, the freezing of the electronic moments induces a very fast decrease of the asymmetry on a $0.1 \mu$s range while the long-time 1/3$^{\rm rd}$ tail monitors the relaxation induced by the spin dynamics.

\begin{figure}
	\includegraphics[width=\columnwidth]{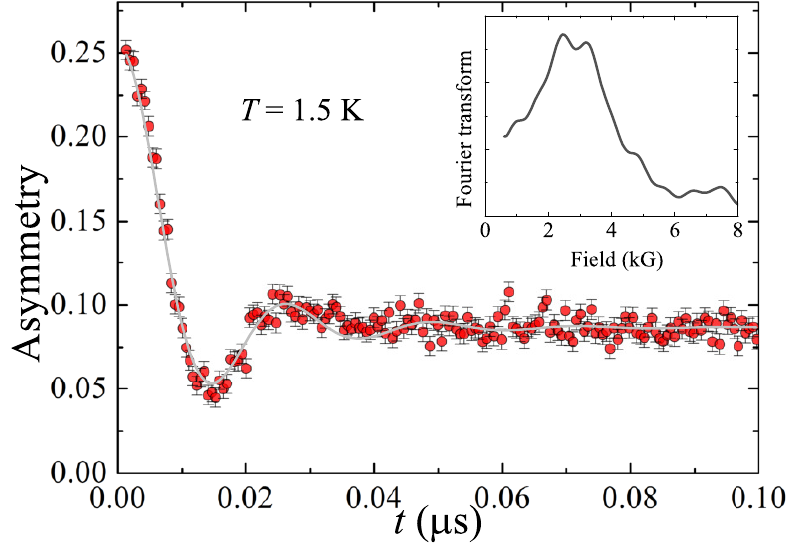}
	\caption{\label{low-T} Early-time $\mu$SR asymmetry in zero external
		field at 1.5~K and fit with a zero-order Bessel function $A[\frac{2}{3} J_0(t) e^{-\lambda t}+1/3]$. Inset: Corresponding Fourier transform after subtraction of the 1/3$^{\rm rd}$ tail.}
\end{figure}

We first focus on the data obtained deep in the frozen regime. Figure~\ref{low-T} shows the very early time behavior of the asymmetry at 1.5~K. The oscillating behavior is made clear by the two visible wiggles and rules out a spinglass-like random freezing. The damping is very large, associated with a large distribution of the field at the muon site as depicted by the Fourier transform presented in the inset. Finally, one also observes the beginning of the 1/3$^{\rm rd}$ tail ($t>0.06\ \mu$s).

The $T=1.5$~K asymmetry evolution with time was fitted according to the model:
\begin{equation}
A(t)=A_0 \left[\frac{2}{3} P_{\rm osc}(t)e^{-({\sigma'}^2 t^2/2)}+\frac{1}{3}\right]e^{-(t/T_1)^{\alpha}}+A_b,
\end{equation}
where $P_{\rm osc}(t)$ is an oscillating function corresponding to the $\mu^+$ precession around the internal magnetic field ($B_{\mu}$) direction at the average frequency $\nu=(\gamma_{\mu}/2\pi)B_{\mu}$, while the damping of the oscillation is due to the width of the internal field distribution $\sigma '/\gamma_{\mu}$.
The constant background, $A_b=0.01$, was fixed from the remaining wTF asymmetry at $T \ll T_{\rm N}$ and $A_0$ was set to $A_0=A_{\rm tot}-A_b$, with $A_{\rm tot}=0.255$, as determined from the total asymmetry fitted above $T_{\rm N}$. We found a slightly better $\chi ^2$ when fitting the oscillations with a Bessel function rather than a simple cosine.
A $B_\mu\sim 2.75$~kG internal field is estimated from the frequency of the oscillations. The fast early time damping reveals a large 1.4~kG HWHM distribution of $B_\mu$, either associated with an incommensurate order (Bessel fit) or a sizeable disorder (cosine fit). The flatness of the 1/3$^{\rm rd}$ tail on the timescale of Fig.~\ref{low-T}, hence the weakness of the relaxation rate $1/T_1$, indicates a purely static magnetic frozen phase in the low-$T$ limit.

We now focus on the relaxation effects evidenced on the 1/3$^{\rm rd}$ tail at $t > 0.1-0.2~\mu$s as displayed in the main panel of Fig.~\ref{ZFmuSR}. This was the central focus of our $\mu$SR study, given the wipe-out observed in NMR. For each temperature below 19.9~K, we could estimate the "unfrozen" contribution so that the asymmetry of the corresponding 1/3$^{\rm rd}$ tail was fixed from the wTF data. The asymmetry evolution with time was fitted using the following functions for $t>t_{1/3^{\rm rd}}$ taken such that $A(t)\leq A_{\rm tail}$~:
	\begin{equation}
		A(t)=A_{\rm tail}\ e^{-{t/T_1}^\alpha} + A_b \quad (T< T_N)
\end{equation}
and
\begin{equation}
	A(t)=A_{0}\ {\rm KT}(t)\ e^{-{t/T_1}^\alpha} + A_b \quad (T \gtrsim T_N),
\end{equation}
where KT($t$) is the Kubo-Toyabe function, Gausian like in the ${t\rightarrow 0}$ limit and $t_{1/3rd}$ is the time where the asymmetry falls to the value expected for the 1/3rd tail, determined as $A_{\rm tail} + A_b$, see above. The stretched exponent $\alpha$ was found to vary smoothly between 12~K and up to the transition from its low-$T$ value $\sim 0.3$ to 1. This indicates a broad distribution of relaxation times, in line with the distribution of internal fields reported in Fig.~\ref{low-T}. The error bars mainly come from the estimate of $t>t_{1/3^{\rm rd}}$, the error bars on the background and on the total asymmetry which slightly impact the relaxation rate. The fits were straightforward between 1.5 and 17.3~K where the "unfrozen fraction" is zero or marginal. For $T = 19.9$~K, the unfrozen part has a contribution to the relaxation which cannot be disentangled from the 1/3rd tail. $A_{\rm tail}$ was therefore replaced by $A_{\rm tail} + A_{\rm osc}$ in Eq.~(11), leading to an average value of the relaxation rate. The increase of the relaxation rate is clearly visible, up to $T_1^{-1}$(15~K)$\sim$1~$\mu$s$^{-1}$, followed by a decrease at $T=17$~K (Fig.~\ref{wTF}, left). The large $T$-range below $T_{\rm N}$ where slow fluctuations occur, clearly signals an unconventional dynamical regime peaked around $0.7~T_{\rm N}$, at variance with the common phase transitions where the relaxation is peaked at $T_{\rm N}$ ~\footnote{Note that the ratio of the NMR and $\mu$ SR relaxation rates $(1/T_1)_{\rm NMR}/(1/T_1)_{\mu{\rm SR}}$ scales with the ratio of the coupling constants squared. An order of magnitude can be calculated from the ratio of $\mu$SR internal field found at 1.5~K and the product of the hyperfine constant by the Cr$^{3+}$ moment, assumed to be 3.87$\mu_{\rm B}$ or accordingly from the 5~kG HWHM  of the NMR spectrum at 1.5~K. At 15~K, this yields a rough estimate of $(T_1)_{\rm NMR}$, of the order of 9-25 $\mu$s, which explains well the wipe-out.}.


\subsection{Microscopic Analysis}
\begin{table}
	\caption{\label{tab:exchange}
		Exchange couplings $J_i$ (in\,K) and single-ion anisotropy $D$ (in\,K) obtained from DFT+$U$ calculations for the spin Hamiltonian, Eq.~\eqref{eq1}. $J_1$, $J_2$, and $J_3$ are in-plane couplings between first, second, and third neighbors, respectively. $J_{\perp}$ is the frustrated interplane coupling.}
	\begin{ruledtabular}
		\begin{tabular}{c@{\hspace{2em}}ccccc}
                             	& $J_1$ & $J_2$ & $J_3$ & $J_{\perp}$ & $D$ \\\hline
LiCrO$_2$                     &  84   & $-0.3$& 3.6   & 0.4         & 0.4 \\
NaCrO$_2$                     &  48   & $-0.5$& 2.7   & 0.1         & 0.6 \\
$\alpha$-KCrO$_2$             &  15   & 0.6   & 4.1   & 0.2         & 0.9 \\
$\alpha$-HCrO$_2$, $R\bar 3m$ &  33   & $-0.2$& 4.3   & 2.8         & 0.5 \\
$\alpha$-HCrO$_2$, $R3m$      &  28   & 1.0   & 4.3   & 2.0         & 0.5 \\
   \end{tabular}
	\end{ruledtabular}
\end{table}
DFT calculations are used to identify the trends in the microscopic magnetic parameters across the $A$CrO$_2$ series. Exchange couplings and single-ion anisotropies computed for LiCrO$_2$, NaCrO$_2$, $\alpha$-KCrO$_2$, and for the two structural models of HCrO$_2$ are summarized in Table~\ref{tab:exchange}. All compounds are dominated by the nearest-neighbor in-plane coupling $J_1$. This coupling is highly sensitive to the Cr--Cr distance $d$ and decreases from LiCrO$_2$ ($d=2.901$\,\r A) to NaCrO$_2$ ($d=2.975$\,\r A) and eventually $\alpha$-KCrO$_2$ ($d=3.044$\,\r A). The $J_1$ value in $\alpha$-HCrO$_2$ lies in between those of the Na and K compounds, despite its Cr--Cr distance of 2.968\,\r A, which is slightly shorter than in the Na case. This additional reduction in the $J_1$ value may be caused by the hydrogen atoms that are placed next to oxygen and change its polarization~\cite{lebernegg2013}. The exact position of hydrogen along the O--H--O contact ($R\bar 3m$ vs $R3m$ models) plays only a minor role, as evident from Table~\ref{tab:exchange}.

The computed values of $J_1$ are generally in good agreement with the experimental estimates based on the magnetic susceptibility data (Table~\ref{table1}). The coupling in LiCrO$_2$ is also in accord with the spectroscopic measurements of magnetic excitations~\cite{Toth13547} that suggested $J_1\simeq 70$\,K (6.0\,meV).

A weak easy-axis single-ion anisotropy ($D$) is present in all compounds and does not change significantly across the series. Indeed, the calculated value of $D$ for $\alpha$-KCrO$_2$ matches well with the value estimated from the ESR experiments (see, Supplementary Materials). On the other hand, the interlayer coupling $J_{\perp}$ notably increases in $\alpha$-HCrO$_2$, owing to the reduced interlayer spacing (Table~\ref{table1}) and the covalent nature of the O--H bonds as opposed to the purely ionic bonding between alkali metals (Li, Na, K) and oxygen.

\begin{table}
	\caption{\label{table1} Comparison of different structural and magnetic parameters
		of $A$CrO$_{2}$ ($A=$ H, Li, Na, and K)~\cite{Angelov213}. The exchange coupling $J~(= J_1)$ for Li, Na, and K compounds is deduced using the HTSE given by Delmas et al~\cite{Doumerc745} whereas for $\alpha$-HCrO$_2$, we have used 11$^{th}$ order HTSE [Eq.~\eqref{eq5}] suggested by Schmidt et al~\cite{Schmidt104443}.}
	\begin{ruledtabular}
		\begin{tabular}{ccccc}
			Compounds & Interplanar & $J$~(K) & $T_{\rm N}$~(K) & Refs.\\
			          & spacing (\AA) &  &  &  \\ \hline
			$\alpha$-HCrO$_{2}$ & 4.774 & 24 & 22.5 & this work \\
			LiCrO$_{2}$ & 4.807 & 78 & 62 & \cite{Hemmida054406,Xiao180401}\\
			NaCrO$_{2}$ & 5.323 & 40 & 41  & \cite{Hemmida054406,Olariu167203} \\
			$\alpha$-KCrO$_{2}$ & 5.963 & 24 &  23  & \cite{Zafar241,Xiao180401} \\
		\end{tabular}
	\end{ruledtabular}
\end{table}

\section{Discussion}
\begin{figure}
	\includegraphics[width=\columnwidth] {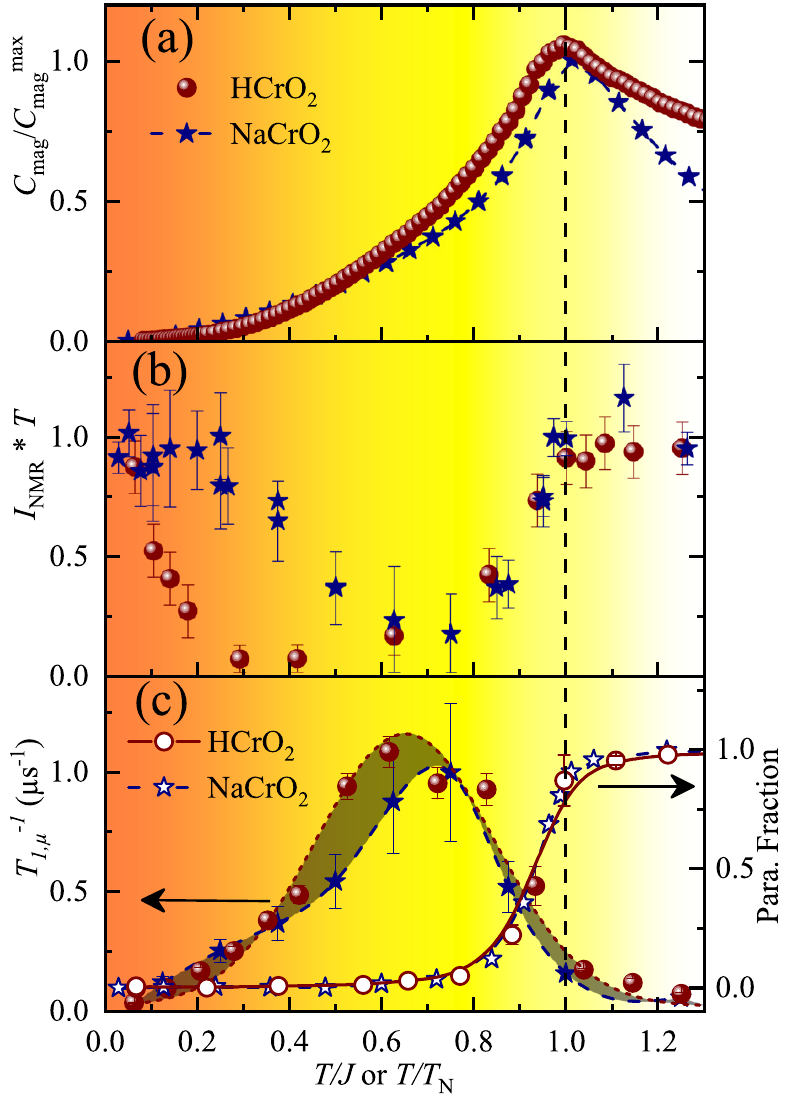}
	\caption{\label{Fig11} Comparison of low temperature data of $\alpha$-HCrO$_2$ with NaCrO$_2$; (a) magnetic heat capacity normalized to its maximum value vs $T/J$ (b) evaluation of integrated NMR intensity with $T/J$, and (c) $\mu$SR relaxation rate $1/T_{1,\mu}$ vs $T/J$ and the paramagnetic fraction vs $T/T_{\rm N}$ in the left and right $y$-axes, respectively. Here the transition temperature is taken to be $T_{\rm N} \simeq 22.5$~K and 41~K for $\alpha$-HCrO$_2$ and NaCrO$_2$, respectively.}
\end{figure}
This discussion section is divided into two parts: the first part is devoted to a discussion on the transition temperature with respect to frustration. The second part is devoted to the extended dynamical regime below the transition which is found to have a universal character among the $A$CrO$_2$ ($A$=H, Na, and K) series of chromates.

For a purely 2D triangular Heisenberg antiferromagnet, one would expect a transition only at $T=0$ into a three-sublattice structure with spins at 120$^{\circ}$. In the presence of a weak interlayer exchange coupling, the transition is still mainly driven by the growth of the 2D correlation length $\xi (T)$ when cooling down: it diverges at $T=0$. Indeed, in a mean-field approach, the transition is obtained by equating the thermal energy to the interaction of 2D correlated patches through a small interlayer exchange coupling. This can be approached by the following mean-field equation that leads to a finite transition temperature,
\begin{equation}
\label{eq13}
	k_{\rm B} T_{\rm N}\sim J_\perp S^2 [\xi ( T_{\rm N})/a]^2.
	\end{equation}
From Refs.~\cite{Azaria1762,Elstner1629} we get
\begin{equation}
\label{eq14}
\xi (T) \sim (T/J)^{-0.5}\exp (\alpha J/T).
\end{equation}
One can then expect a transition temperature modestly increasing with $J_\perp$. From Tables~I and II, one can notice that $J_\perp$ decreases when the inter-layer distance increases from $\alpha$-HCrO$_2$ to $\alpha$-KCrO$_2$, as expected, and $J_\perp$ is more than a factor ~15-30 larger for $\alpha$-HCrO$_2$ as compared to NaCrO$_2$ and $\alpha$-KCrO$_2$. 
While the $J/T_{\rm N}$ ratio is similar or even slightly larger for $\alpha$-HCrO$_2$, the transition temperature surprisingly does not scale at all with $J_\perp$. We suggest that this failure of the mean field, well-established, approach for low-dimensional systems can be assigned to a peculiar degree of interlayer frustration in $\alpha$-HCrO$_2$. Indeed, we first note that the geometry of the inter-layer coupling [see Fig.~\ref{Fig1}(b)] leads to some frustration through the {\it tetrahedral} units resulting from the $ABC$ type of stacking, common to all the chromates mentioned above: one Cr$^{3+}$ is coupled to six Cr$^{3+}$ from the two adjacent layers. Second, a higher degree of frustration induced by the interlayer coupling, as found for $\alpha$-HCrO$_2$, certainly prevents the correlation length to grow according to Eq.~\eqref{eq14} which leads to the apparent paradox of "a larger $J_\perp$ yields to a lower $T_{\rm N}$". 
We note that this frustration scenario is in line with the incommensurate structure observed at 1.57~K through the NMR lineshape and also consistent with the $\mu$SR time-evolution of the asymmetry at 1.5~K.

Now, coming to the intermediate $T$-range below $T_{\rm N}$, several of the $A$CrO$_2$ compounds show a very similar phenomenology. In Fig.~\ref{Fig11}, we compare our results to those obtained in NaCrO$_2$~\cite{Olariu167203}, using $J$ as a scaling parameter. \\
({\it i}) Both $\alpha$-HCrO$_2$ and NaCrO$_2$ undergo a magnetic transition that manifests itself in the specific heat and $\mu$SR, but does not lead to the formation of a conventional 3D transition with a narrow critical regime centered in the vicinity of $T_{\rm N}$. On the contrary, here the transition at $T_{\rm N}$ is followed by a broad regime with slow fluctuations peaked around 0.6-0.7$~T_{\rm N}$ [Fig.~\ref{Fig11}(c)]. Strikingly, this maximum largely coincides with the bifurcation point of the FC/ZFC susceptibility reported at 12-15~K in Fig.~\ref{Fig3} for $\alpha$-HCrO$_2$. The extension of this regime can be tracked through the $\mu$SR relaxation rate, which shows a striking similarity in both compounds once the temperatures are normalized by $J$.\\
({\it ii}) The heat capacity also displays a very broad maximum around $T_{\rm N}$ and a similar behavior below $T_{\rm N}$ for both compounds when the temperature axis is scaled by their respective $J$.\\
({\it iii}) For both compounds, the intensity of the NMR line progressively decreases, goes through a minimum, and recovers below a temperature $T^*$ where slow fluctuations detected through $\mu$SR freeze out~\footnote{Note that the difference in the extension of this regime can be due in the way extrapolations of transverse relaxation $T_2$ plots have been performed which is always fairly delicate when $T_2$ is short.}.

Similar plots including the data taken on $\alpha$-KCrO$_2$ can be found in the supplementary material section, they clearly demonstrate the universal character of this scaling in $T/J$ for that series of chromates.

We are therefore led to the conclusion that a broad universal dynamical regime emerges on a $T/J$ scale, typical of $ABC$-stacked triangular Heisenberg antiferromagnets and characterized by a very progressive slowing down of spin fluctuations.

Neutron scattering studies offered a possible explanation for this unusual behavior as reported earlier for NaCrO$_2$~\cite{Hsieh3174,Hsieh1341,Hsieh2014}. Between 1.3 and 0.7~$T_{\rm N}$, the in-plane correlation length grows rapidly with a $\vec{q}$ vector typical of the $120^{\circ}$ pattern. On the contrary, the inter-layer correlation length increases very progressively from 1 inter-layer spacing at $T_{\rm N}$ to only 4 inter-layer spacings at 0.7~$T_{\rm N}$ where it levels off. This $T$-range closely below $T_{\rm N}$ is dominated by 2D correlations.

The increase in $1/T_{1,\mu}$ up to its maximum could be such a hallmark of the 2D regime, in fair agreement with the $T^2$ behavior of the specific heat that we clearly observe in $\alpha$-HCrO$_2$. At lower temperatures, a release of the inter-layer frustration through a small incommensuration may strengthen the 3D coupling. The consequence is a progressive slowing down of the fluctuations as observed in $\mu$SR. This regime is typical of a 2D-3D crossover. What remains unclear is certainly the universality observed in the maximum of $1/T_{1,\mu}$ . It might be a coincidental result of two counteracting mechanisms driven by $J_\perp$, namely, an increase of both the coupling and the frustration between the layers.

Alternative scenarios, namely, the Berezinskii-Kosterlitz-Thouless (BKT) scenario of vortex-antivortex binding or that of $Z_2$ vortex excitations, have been proposed for NaCrO$_2$~\cite{Olariu167203,Hemmida054406}. Indeed the maximum in the $\mu$SR relaxation rate at $T_m\simeq 0.7\; T_{\rm N}$ was found to coincide with the divergence of the correlation length inferred from the broadening of the ESR line according to the BKT model, $\xi(T) \sim \exp{\frac{b}{T-T_m}} $. The data now available on $\alpha$-HCrO$_2$ and $\alpha$-KCrO$_2$ prove the failure of this scaling in $T/T_m$ (see the Supplementary Materials for the tentative scaling) and rules out such an interpretation. In addition, we note that a BKT scenario requires a substantial XY anisotropy, which has never been observed experimentally. This, therefore, calls for a different interpretation of the ESR line broadening in terms of a growth of the correlation length when entering the 3D regime.

A more general lesson from the $A$CrO$_2$ chromates is that the fluctuating 2D regime clearly sets in below $T_{\rm N}$, likely from the interlayer frustration. This offers a new perspective on the persistent spin dynamics and other spin-liquid phenomenology such as observed in the $A$Yb$X_2$ compounds~\cite{ding2019,bordelon2019} where exchange anisotropy also certainly plays a role. It also calls for the improved theoretical understanding of how deep the frustration of weak interlayer couplings may affect the ground state of a quasi-2D antiferromagnet. Our experimental data for the pure Heisenberg triangular chromates set an important benchmark for such studies.


In conclusion, besides the observation of a N\'eel transition, features reminiscent of the pure 2D Heisenberg triangular lattice are manifested through the dynamics as revealed here experimentally, between $T_{\rm N}$ and $0.7\; T_{\rm N}$. The study of the $A$CrO$_2$ chromates ($A$ = H, Na, and K) points to an original and universal character of excitations. With their half filled $t_{2g}$ orbitals, chromates are certainly the best representative of a semi-classical $S=3/2$ Heisenberg model. The perfection of the equilateral triangular lattice and the absence of disorder are strong assets to provide a solid playground for the physics of frustration at play on the triangular lattice in a context where more disordered compounds such as YbMgGaO$_4$ or $A$Yb$X_2$ QSL candidates are in the spotlight. Certainly, the interpretation of the 2D-3D crossover resides in the details of each of these compounds although the universal plot in $T/J$ might suggest that it is still governed by the dynamics specific to the 2D frustrated character of the triangular lattice.

\section{Summary}
The static and dynamic properties of spin-$3/2$ TLHAF $\alpha$-HCrO$_2$ are studied in detail and compared with the iso-structural compounds (Na,K)CrO$_2$. $\chi(T)$ could be modeled using HTSE for spin-$3/2$ TLHAF which yields an intra-layer coupling $J/k_{\rm B} \simeq 24$~K. The complementary band structure calculations result in a similar value of $J$. They additionally provide a sizable inter-layer coupling $J_{\perp}/k_{\rm B} \simeq 2.8$~K which appears specific to $\alpha$-HCrO$_2$. This brings in inter-layer frustration which shifts down the transition temperature as compared to other members of the chromates series. On the experimental side, a large hyperfine coupling $A_{\rm hf} \simeq 1021$~Oe/$\mu_{\rm B}$ between the $^{1}$H nuclei and the Cr$^{3+}$ spins corroborate the existence of such a sizable inter-layer coupling.

Although, $\chi(T)$, $C_{\rm p}(T)$ and muon asymmetry measurements reveal the onset of a magnetic transition at $T_{\rm N}$, which is also monitored through a magnetoelastic coupling, a wide fluctuating crossover regime marked by a broad peak in the muon relaxation rate $1/T_{1,\mu}$ centered at $T \simeq 0.7~T_{\rm N}$ and a minimum in the $^1$H NMR integrated intensity is singled out. Apparently, this slow dynamical regime turns out to be a universal character of the TLHAFs $A$CrO$_2$ ($A=$ Na, K, and H) when the specific heat, NMR intensity, $1/T_{1,\mu}$, and paramagnetic fraction are plotted against the scaled (with respect to $J$ or $T_{\rm N}$) temperature. This supports a scenario where a crossover from 2D to 3D correlations sets in around $0.7~T_{\rm N}$ preceded by a typical 2D regime of the TLHAF.

\acknowledgments
PM and RN thank J. Richter for discussions and for providing the HTSE and Pad\'{e} approximants used in the fit of our susceptibility data. KS and RN would like to acknowledge SERB, India for financial support bearing sanction Grant No. CRG/2019/000960. The work in Augsburg was supported by the German Research Foundation (DFG) via the Project No. 107745057 (TRR80). Work at the Ames Laboratory was supported by the U.S. Department of Energy, Office of Science, Basic Energy Sciences, Materials Sciences, and Engineering Division. The Ames Laboratory is operated for the U.S. Department of Energy by Iowa State University under Contract No. DEAC02-07CH11358. FB, PM, and GS acknowledge the support of the French Agence Nationale de la Recherche under Grant No. ANR-18-CE30-0022. The work of GS is funded by the Swiss National Science Foundation Mobility grant P2EZP2-178604 and PALM LabEx grant ANR-10-LABX-0039-PALM. Part of this work is based on experiments performed at the Swiss Muon Source S$\mu$S, Paul Scherrer Institute, Villigen, Switzerland. We would like to thank A. Amato and C. Wang for the technical assistance with the GPS spectrometer. We thank C. Delmas (ICMCB, Bordeaux) for providing the KCrO$_2$ sample and N. Penin for mounting it. We thank A. Ozarowski for technical assistance with ESR measurements. AZ acknowledges the financial support of the Slovenian Research Agency through program No.~P1-0125 and projects No.~BI-US/18-20-064, No.~J1-2461, and No.~N1-0148. The National High Magnetic Field Laboratory where the ESR investigation was conducted is supported by National Science Foundation through NSF/DMR-1644779 and the State of Florida.

%

\widetext
\clearpage
\begin{center}
	\textbf{\large Supplementary Materials}
\end{center}
\begin{center}
	\textbf{\large "Universal fluctuating regime in triangular chromate antiferromagnets"}
\end{center}
\setcounter{equation}{0}
\setcounter{figure}{0}
\setcounter{table}{0}
\setcounter{page}{1}
\makeatletter
\setcounter{section}{0}
\renewcommand{\thesection}{S-\Roman{section}}
\renewcommand{\thetable}{S\arabic{table}}
\renewcommand{\theequation}{S\arabic{equation}}
\renewcommand{\thefigure}{S\arabic{figure}}
\renewcommand{\bibnumfmt}[1]{[S#1]}
\renewcommand{\citenumfont}[1]{S#1}

\section{AC Susceptibility}
\begin{figure}[h]
	\includegraphics[] {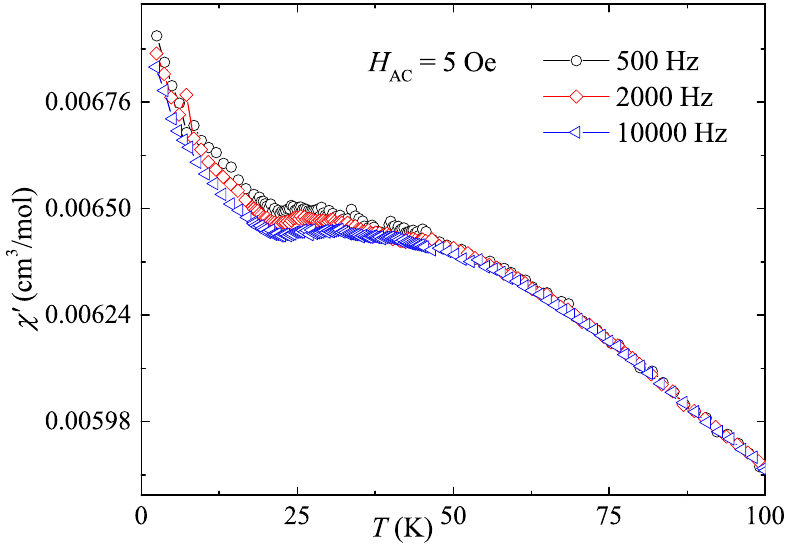}
	\caption{\label{S1} Real part of the ac susceptibility ($\chi^{\prime}$) of $\alpha$-HCrO$_2$ vs $T$ measured in an ac field of $H_{\rm AC} = 5$~Oe and at three different frequencies.}
\end{figure}
The real part of the ac susceptibility ($\chi^{\prime}$) as a function of temperature is shown in Fig.~\ref{S1}. A weak anomaly is observed at $T_{\rm N}$ which is frequency independent. No features are visible around 12~K where ZFC and FC susceptibilities show bifurcation, ruling out the possibility of spin-glass transition.

\section{NMR Relaxation Rates}
\begin{figure}[h]
	\includegraphics[] {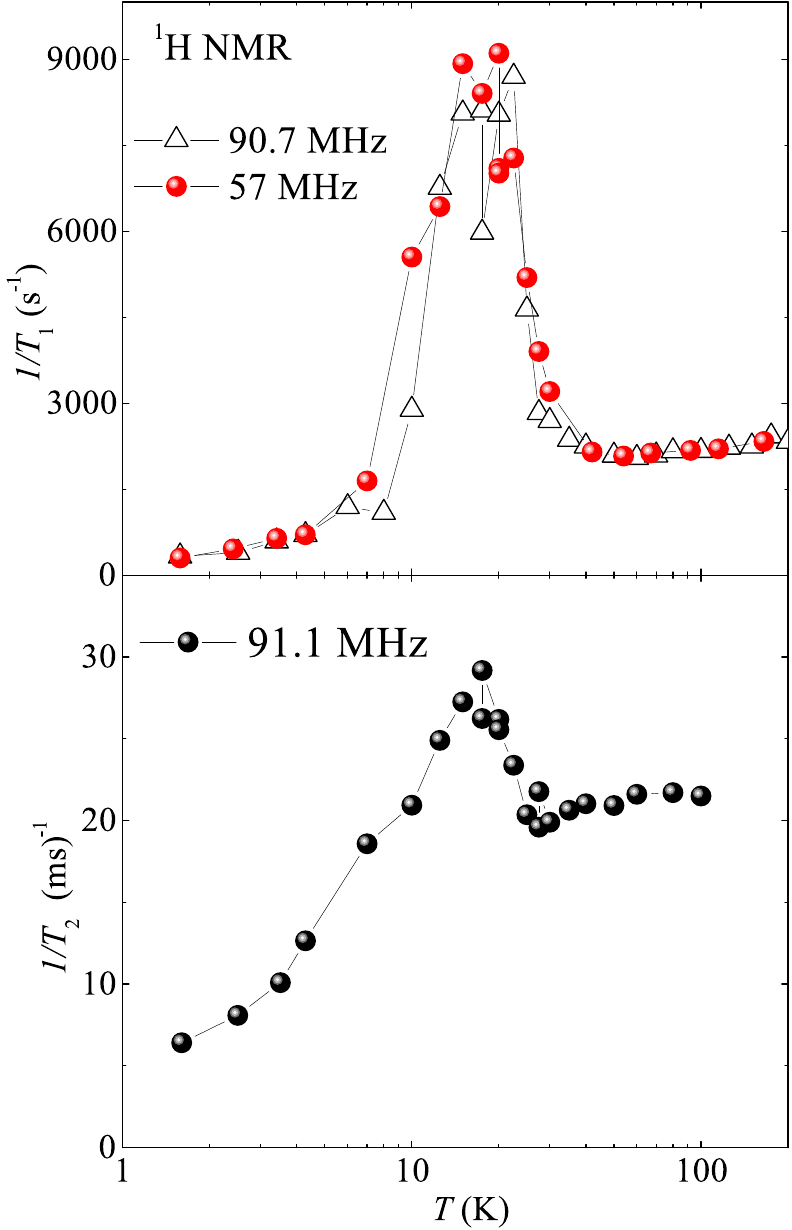}
	\caption{\label{S2} Upper panel: Temperature dependence of the $^{1}$H spin-lattice relaxation rate ($1/T_{1}$) of $\alpha$-HCrO$_2$, measured at two different frequencies. Lower panel: Temperature variation of the $^{1}$H spin-spin relaxation rate ($1/T_2$) for $\alpha$-HCrO$_2$.}
\end{figure}
$^{1}$H nuclear spin-lattice relaxation rate $1/T_{1}$ was measured at the central position of the NMR spectral line.
For a $I=1/2$ nucleus, the recovery of the longitudinal magnetization is expected to follow a single-exponential behavior. In $\alpha$-HCrO$_{2}$, the recovery of the nuclear magnetization after a saturation pulse was indeed fitted well by the stretch exponential function
\begin{equation}
	1-\frac{M(t)}{M_{0}}=Ae^{-(t/T_{1})^{\eta}},
	\label{eq12}
\end{equation}
where $M(t)$ is the nuclear magnetization at a time $t$ after the saturation pulse, $M_{0}$ is the equilibrium magnetization, and $\eta$ is the exponent. The $\eta$ parameter is close to 1 for $T>T_{\rm N}$ suggesting a single relaxation time that confirms high homogeneity of the sample and its magnetism. $\eta$ values below 1 were obtained at lower temperatures.

Temperature evolution of $1/T_1$ does not show any appreciable frequency dependence (upper panel of Fig.~\ref{S2}). $1/T_1$ diverges around $T_{\rm N}$ and develops a broad hump between $\sim 10$ and 30~K. At high temperatures ($T \gtrsim 30$~K), $1/T_{1}$ is almost temperature-independent, typical in the high-temperature limit ($T\gg J/k_{\rm B}$) because spins are uncorrelated~\cite{Moriya23s}.

The evolution of $1/T_1$ at lower temperature is remarkable. Conventional antiferromagnets show a sharp peak at the magnetic ordering temperature followed by a fast decay towards low temperatures caused by the suppression of magnon excitations~\cite{Beeman359s}. Such a decay is indeed observed in $\alpha$-HCrO$_2$, but only beyond a broad hump developing between 10 and 30~K. Similar features have been seen in some other frustrated antiferromagnets, where they are ascribed to the abundance of low-energy excitations~\cite{Majumder214417s}. However, those systems develop static magnetic order below $T_{\rm N}$, which is not the case in $\alpha$-HCrO$_2$, as we show here. Moreover, because of much shorter time window, $\mu$SR is more suitable than NMR for probing the slow dynamics, especially in the fluctuating regime. In such a case, NMR can only detect a fraction of the total nuclei with longer relaxation.

To measure the relaxation of transverse nuclear magnetization $M_{xy}(2t)$, a spin-echo sequence was used with varying time ($t$) between the pulses. The recovery curves were well fitted by an exponential function
\begin{equation}
	M_{xy}(2t) = M_{0}e^{-2t/T_{2}}.
	\label{eq15}
\end{equation}
The obtained spin-spin relaxation rate $1/T_2$ data are plotted as a function temperature in the lower panel of Fig.~\ref{S2}. It also shows a very broad peak (10-30~K) with the center of gravity at $T \simeq 17$~K, similar to the $1/T_1$ data.
\newpage
\section{Universal behavior of chromates $A$CrO$_2$, $A$ = H, Na, and K}
\begin{figure}[h]
	\includegraphics[width=12cm] {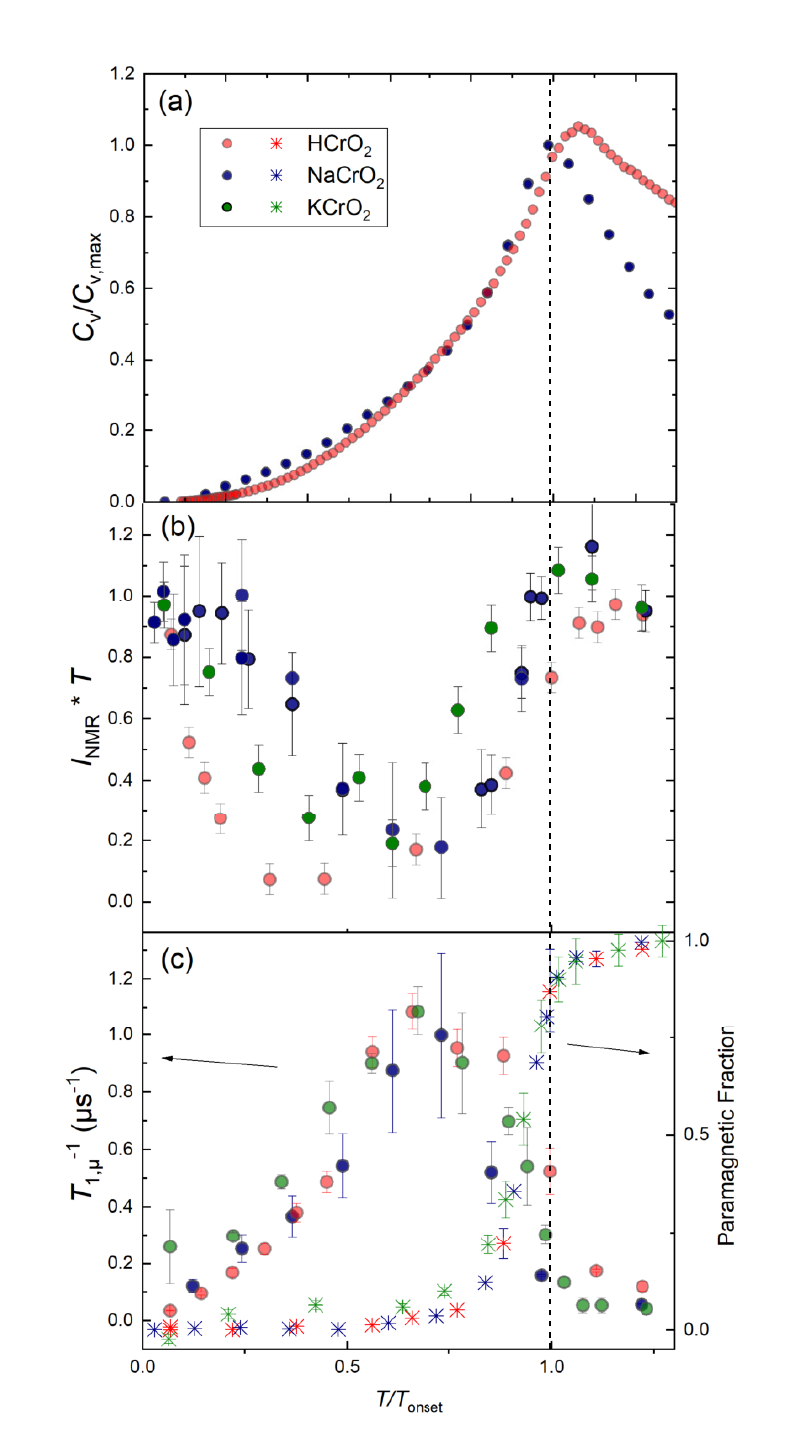}
	\caption{\label{S3} Plot of low temperature data of $A$CrO$_2$, $A$ = H, Na, and K; (a) normalized magnetic heat capacity to its maximum value vs temperature for $\alpha$-HCrO$_2$ and NaCrO$_2$, (b) evaluation of integrated NMR intensity with temperature, and (c) temperature variation of $\mu$SR relaxation rate $1/T_{1,\mu}$ and the paramagnetic fraction. For the latter and for the NMR intensity, a 21\% impurity phase has been subtracted. The temperature in the $x$-axis is normalized by $T_{\rm N} \simeq 22.5$~K, 41~K, and 24.6~K for $\alpha$-HCrO$_2$, NaCrO$_2$, and $\alpha$-KCrO$_2$, respectively.}
\end{figure}
Similar $\mu$SR (PSI facility), $^{19}$K NMR experiments were performed for $\alpha$-KCrO$_2$ along those for $\alpha$-HCrO$_2$. Given its high reactivity, the sample which had been synthesized in the 70's and kept in sealed ampoules since then was transferred under He atmosphere into air tight sample holders suited for each experiment and filled with He gas to ensure a good thermalization. A 21\% part of the sample was found to be degraded both from $^{19}$K NMR and $\mu$SR. The contribution from that part gives clear-cut signals and can safely be separated from the relevant one in all three experiments.\\
Similar to Fig.~\ref{Fig11}, a comparison of our $\mu$SR and NMR results for the three compounds is presented in Fig.~\ref{S3}, as plotted versus $T/T_{\rm N}$. Clearly, the universal character of the $T/T_{\rm N}$ scaling is confirmed.

\section{Failure of the BKT scenario}
\begin{figure}[h]
	\includegraphics[width=12cm] {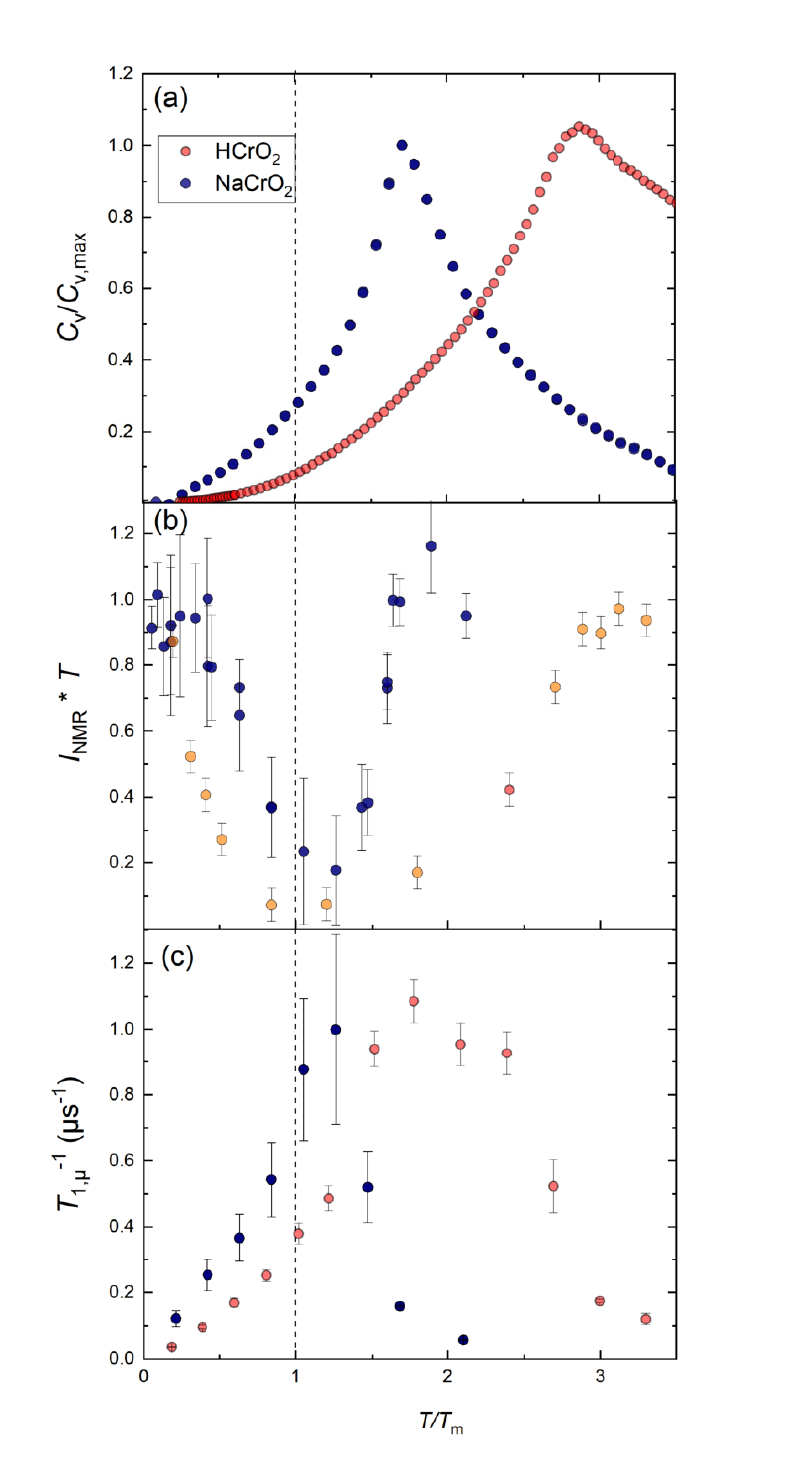}
	\caption{\label{S4} Same plot as in Fig.~\ref{Fig11}. The temperature in the $x$-axis is normalized by the presumed BKT transition temperature as reported in Ref.~\cite{Hemmida054406s} $T_{\rm m}/ T_{\rm N} = 0.37$ for $\alpha$-HCrO$_2$ and 0.56 for NaCrO$_2$, respectively.}
\end{figure}
Electron paramagnetic resonance (EPR) measurements by M. Hemmida \textit{et al.}~\cite{Hemmida054406} suggested the occurrence of Berezinskii-Kosterlitz-Thouless (BKT) transition at temperatures below $T_{\rm N}$. The BKT transition is expected at $T_{\rm m} \simeq 0.95T_{\rm N}$, $T_{\rm m} \simeq 0.56T_{\rm N}$, and $T_{\rm m} \simeq 0.37T_{\rm N}$ for LiCrO$_{2}$, NaCrO$_{2}$, and $\alpha$-HCrO$_{2}$, respectively. It was therefore suggested that the fluctuating regime persists below $T_{\rm N}$ and down to $T_{\rm m}$. In Fig.~\ref{S4}, we plot a summary of our $\mu$SR results similar to Fig.~\ref{Fig11} and compare it to NaCrO$_2$ using a $T/T_{\rm m}$~\cite{Hemmida054406s} rather than a $T/T_{\rm N}$ scale. No commonality is found for this BKT scenario.

\section{Electron Spin Resonance on $\alpha$-KCrO$_2$: a check of DFT+U calculations}
Our ESR investigation on $\alpha$-KCrO$_2$ was conducted on a powder sample sealed in a quartz tube on a custom-made homodyne-detection spectrometer working in transmission mode at the NHMFL, Tallahassee, USA. The measurements were performed in the Faraday configuration at the irradiation frequency of 108.8\,GHz in the temperature range between 50 and 290\,K.
Field modulation with modulation amplitude of 2\,mT was used to enhance signal-to-noise ratio and derivative ESR spectra were consequently recorded.

\begin{figure}[h]
\includegraphics[trim = 0mm 0mm 0mm 0mm, clip, width=0.6\linewidth]{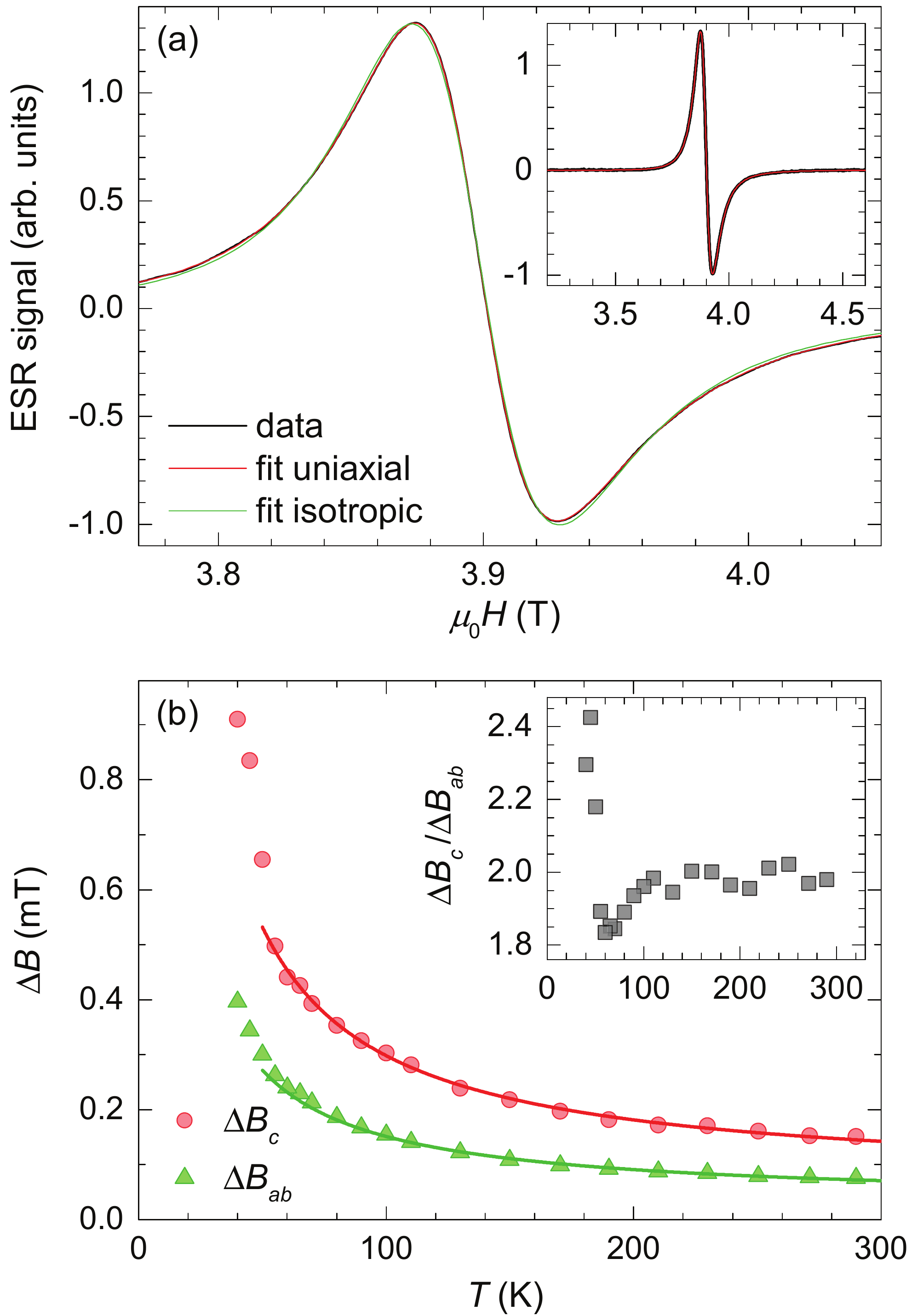}
\caption{(a) The ESR spectrum of $\alpha$-KCrO$_2$ recorded at 290\,K and 108.8\,GHz (black line).
The red and green lines show powder simulations with Lorentzian line shape assuming uniaxial anisotropy ($\chi^2=0.50$) and isotropic symmetry ($\chi^2=4.65$), respectively.
The inset shows a perfect fit of the uniaxial model in a wider field window.
(b) The temperature dependence of the principal ESR line widths for the magnetic field along the crystallographic $c$-axis and within the $ab$-plane.
The solid lines are fits of the two datasets with the model $\Delta B_i = \Delta B_i^\infty + C_i/T$ above 100\,K. The inset shows the temperature variation of the ratio of the two principal line widths.}
\label{S5}
\end{figure}

A typical ESR spectrum recorded at 290\,K is shown in Fig.~\ref{S5}(a).
It is of Lorentzian shape, as typically observed in strongly exchange-coupled spins systems \cite{pake1962paramagnetic}.
The fit with an isotropic Lorentzian model is not far from the experimental data ($\chi^2=4.65$), indicating that the $g$-factor anisotropy is small.
However, allowing for uniaxial $g$-factor anisotropy $g(\theta)=(g_c^2 \cos^2\theta + g_{ab}^2 \sin^2\theta)^{1/2}$ and line-width anisotropy $\Delta B(\theta)=(\Delta B_c^2 \cos^2\theta + \Delta B_{ab}^2 \sin^2\theta)^{1/2}$ fits the experiment considerably better ($\chi^2=0.50$).
Here $\theta$ denotes the angle between the high-symmetry crystallographic $c$-axis and the applied magnetic field.
Such uniaxial anisotropy is due to trigonal distortion of CrO$_6$ octahedra -- it reflects a three-fold rotational axis that is parallel to the crystallographic $c$-axis and passes through the Cr site.
The fit with the uniaxial model discloses a small $g$-factor anisotropy [$g_c=1.995(3)$, $g_{ab}=1.989(3)$] and a large line-width anisotropy [$\Delta B_c=151(1)$~mT, $\Delta B_{ab}=76(1)$~mT], yielding $\Delta B_c/\Delta B_{ab}=2.0$.
The corresponding peak-to-peak line width $\Delta B_{pp} = 53$~mT is in good agreement with previously reported values of 49 and 52\,mT, recorded at lower irradiation frequencies in $X$-band and $Q$-band, respectively\cite{angelov1984relation}. 

With decreasing the temperature below 290\,K the ESR spectrum of $\alpha$-KCrO$_2$ progressively broadens and becomes too broad for a reliable fit below $\sim 50$\,K [Fig.~\ref{S5}(b)].
Interestingly, the ratio of the two principal line-widths $\Delta B_c/\Delta B_{ab}=2.0$ found at 290\,K remains unchanged down to $\sim100$\,K, where it starts to decrease with decreasing temperature and then notably increases below $\sim 70$\,K [inset of Fig.~\ref{S5}(b)].
As the temperature dependence of the ESR line width is a sign of developing spin correlations, these thus seem to be isotropic above $\sim$100\,K and become anisotropic below this temperature.
In the regime of isotropic spin correlations both line widths follow the temperature dependence $\Delta B_i(T) = \Delta B_i^\infty + C_i/T$, where the term $\Delta B_i^\infty$ denotes the infinite-temperature limit of the line width and the term $C_i/T$ corresponds to the predicted lowest order correction in $1/T$~\cite{pake1962paramagnetic}.
The limiting line widths $\Delta B_c^\infty=65(1)$~mT and $\Delta B_{ab}^\infty=31(1)$~mT that are found after fitting the line-width datasets above 100\,K are a direct measure of magnetic anisotropy present in $\alpha$-KCrO$_2$.

Within the well-established Kubo-Tomita theory~\cite{kubo1954general}, the ESR full-width-at-half-maximum of an exchange-coupled spin system is determined by the second moment $M_2$ and the fourth moment $M_4$ of the absorption line~\cite{zorko2018determination},
\begin{equation}
\Delta B = \sqrt{2\pi} \frac{k_B}{g \mu _B} \sqrt{\frac{M_2^3}{M_4}},
\label{eq_linewidth}
\end{equation}
with
\begin{align}
M_{2}&= \notag \frac{\left\langle \left[ \mathcal{H}^{\prime },S^{+}\right]
[ S^{-},\mathcal{H}^{\prime }] \right\rangle} {\left\langle S^{+}S^{-}\right\rangle},\\
M_{4}&=\frac{\left\langle \left[ \mathcal{H}-\mathcal{H}_{Z},\left[
\mathcal{H}^{\prime },S^{+}\right] \right] [ \mathcal{H}-\mathcal{H}_{Z},\left[
\mathcal{H}^{\prime },S^{-}\right] ] \right\rangle}{ \left\langle
S^{+}S^{-}\right\rangle},
\label{eq_moments}
\end{align}
where $[A,B]$ denotes a commutator of operators $A$ and $B$, $\left\langle ... \right \rangle$ stands for canonical averaging and $S^+$, $S^-$ are the ladder operators for spin $S=3/2$.
The Hamiltonian of the system $\mathcal{H}=\mathcal{H}_{ex}+\mathcal{H}_Z+\mathcal{H}^{\prime }$ is divided into the exchange term $\mathcal{H}_{ex}$, the Zeeman term $\mathcal{H}_Z$ and an anisotropy term $\mathcal{H}^{\prime }$. The two moments can be calculated analytically at infinite temperature where spin correlations between neighboring sites vanish.

In $\alpha$-KCrO$_2$ the main magnetic anisotropy term is of the single-ion anisotropy type
\begin{equation}
\mathcal{H}^{\prime }=\sum_i D(S_i^z)^2,
\label{eq_aniso}
\end{equation}
due to the crystal-field symmetry being reduced from cubic at the Cr site.
The anisotropy axis is parallel to the anisotropy axis of the $g$ tensor and is thus parallel to the crystallographic $c$ axis.
We note that Dzyaloshinskii-Moriya interaction, often being the dominant magnetic anisotropy term in exchange coupled systems, is absent in $\alpha$-KCrO$_2$ due to a center of inversion in the middle of the Cr-Cr bond.
Inserting magnetic anisotropy given by Eq.\,(\ref{eq_aniso}) into Eqs.\,(\ref{eq_moments}) yields the angular dependence of the ESR line width from Eq.\,(\ref{eq_linewidth}),
\begin{equation}
\Delta B^\infty (\theta) = \sqrt{2\pi} \frac{k_B}{g \mu _B} \frac{6D^2\left(1+\cos^2\theta \right)}{5\sqrt{30}J_1}.
\end{equation}
Here we neglect all exchange interactions beyond the dominant nearest-neighbor exchange interaction $J_1$, because squares of exchange interactions appear in the expression of $M_4$~\cite{zorko2011role}.

The calculated ESR line-width anisotropy yields $\Delta B^\infty_c = 2 \Delta B^\infty_{ab}$, which is in perfect agreement with the experiment above 100\,K [inset in Fig.~\ref{S5}(b)].
The experimentally determined values $\Delta B_c^\infty=65(1)$~mT, $\Delta B_{ab}^\infty=31(1)$~mT and the nearest-neighbor exchange coupling  $J_1=15$\,K from our DFT+$U$ calculations for $\alpha$-KCrO$_2$ (Table\,I in the main text) then predict the single-ion anisotropy $D=1.1(1)$~K, which agrees very well with the theoretical prediction in Table\,I, thus verifying the validity of the theoretical calculations.

\end{document}